\RequirePackage{fix-cm}
\PassOptionsToPackage{table}{xcolor}
\documentclass{article}
\usepackage{preprint,times}
\iclrfinalcopy
\renewcommand{\preprintrunningtitle}{Beyond the Final Embedding: Retrieving Memories from Token-Level Evidence}

\usepackage{amsmath,amsfonts,bm}

\def\eqref#1{equation~\ref{#1}}

\def\1{\bm{1}}

\DeclareMathAlphabet{\mathsfit}{\encodingdefault}{\sfdefault}{m}{sl}
\SetMathAlphabet{\mathsfit}{bold}{\encodingdefault}{\sfdefault}{bx}{n}

\usepackage[T1]{fontenc}
\usepackage{amsmath,amssymb,booktabs,array,tabularx,graphicx,float,placeins}
\usepackage[nopatch=eqnum]{microtype}
\usepackage{needspace}
\usepackage{caption}
\usepackage{xcolor}
\usepackage{hyperref}
\usepackage{url}
\colorlet{oursblue}{black!8}
\makeatletter
\let\tableinput\@@input
\makeatother
\hypersetup{colorlinks=true,linkcolor=paperLink,citecolor=paperLink,urlcolor=paperLink}
\definecolor{paperLink}{RGB}{62,82,112}
\definecolor{groupgray}{RGB}{239,240,241}
\definecolor{gaingreen}{RGB}{36,126,69}
\newcommand{\norm}[1]{\left\lVert #1\right\rVert}

\title{Beyond the Final Embedding: Retrieving\\Memories from Token-Level Evidence}
\author{\textbf{Han Luo$^{1}$ \quad Zhan Zhuang$^{2}$ \quad Xu He$^{2}$ \quad Jujia Zhao$^{3}$}\\
\textbf{Yu Li$^{1}$ \quad Suqin Yuan$^{4}$ \quad Dong Li$^{2}$ \quad Lei Feng$^{1}$\thanks{Corresponding author.}}\\
{\normalfont\normalsize
$^{1}$Southeast University \quad $^{2}$MemoraX AI \quad $^{3}$Leiden University \quad $^{4}$University of Sydney}
}
\hypersetup{
  pdftitle={Beyond the Final Embedding: Retrieving Memories from Token-Level Evidence},
  pdfauthor={Han Luo, Zhan Zhuang, Xu He, Jujia Zhao, Yu Li, Suqin Yuan, Dong Li, Lei Feng},
  pdfsubject={Memory retrieval using token-level evidence}
}

\newcommand{\CostReductionLow}{81.4}
\newcommand{\CostReductionHigh}{85.9}
\newcommand{\SQFourRecallMeanDelta}{+0.33}
\newcommand{\SQFourMeanDelta}{+0.19}

\newcommand{\smallstd}[1]{\kern.1pt\scalebox{0.67}{$\pm$#1}}
\newcommand{\plotcaption}[1]{\caption{#1}}

\begin{document}
\maketitle
\vspace{-20pt}
\suppressfloats[t]

\begin{abstract}
Memory-augmented assistants retrieve evidence from stored conversations and documents to answer queries.
Most embedding-based retrievers compare single-vector representations of queries and memories, even when relevance depends on a small part of the content.
Such local evidence can be matched to queries using the contextual token representations already computed by the same embedding models.
Our method uses sparse token matching to improve retrieval without additional training.
Each query token retrieves a small set of matching tokens across candidate memories.
The matching scores are aggregated for each memory and combined with whole-memory similarity to rank memories.
An optional low-rank query adapter can be trained with supervision while keeping the encoder and stored memory representations fixed.
Our method improves retrieval across eight benchmarks and multiple model families without additional training.
Query adaptation further increases the gains to 13.3 and 14.3 percentage points in average Recall@10 and NDCG@10 on Qwen3-Embedding-4B.
These retrieval gains translate into higher question-answering accuracy on long-term memory benchmarks.
\end{abstract}

\section{Introduction}
Memory retrieval enables assistants to answer questions using information from past conversations and stored documents \citep{maharana2024locomo,wu2025longmemeval,zhao2026lmeb}. Existing retrieval approaches range from lexical methods that rely on word overlap \citep{robertson2009bm25} to semantic methods that compare learned representations of queries and memories \citep{wang2022e5}. Among these, single-vector retrieval encodes each query and memory into a single embedding and uses embedding similarity to rank memories, supporting efficient indexing and search over large memory banks \citep{karpukhin2020dpr,zhang2025qwen}. The expressive limitations of single-vector retrieval have attracted both theoretical and empirical study \citep{weller2026limitations}.

\begin{figure}[t]
\centering\includegraphics[width=\linewidth]{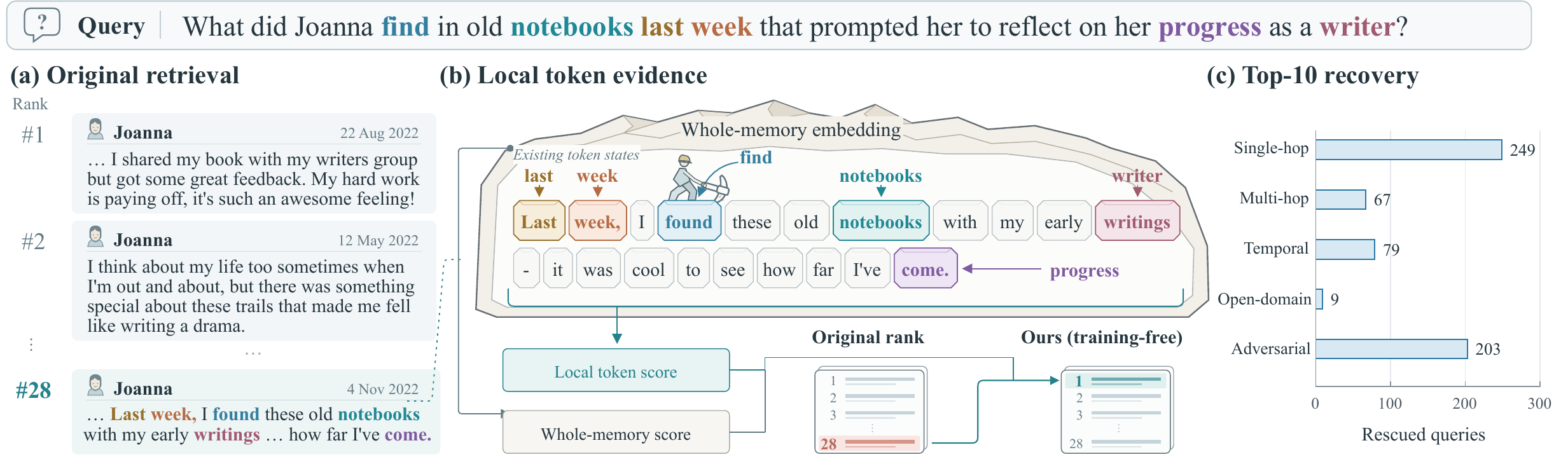}
\caption{\textbf{Local evidence from existing token representations.} Training-free fusion raises the relevant memory from rank 28 to 1 and yields a net gain of 523 Top-10 hits on LoCoMo.}
\label{fig:motivation}
\end{figure}

In practice, the information needed to answer a question may occupy only a small part of a memory, and different questions may depend on different parts. Single-vector retrieval scores relevance through whole-memory similarity without explicitly matching queries to fine-grained information within each memory. However, the same encoder already computes contextual token representations that can be used to match queries to such local evidence. Figure~\ref{fig:motivation} illustrates how combining the resulting local scores with whole-memory similarity improves retrieval. 

These local matches illustrate the use of richer matching signals, a direction that existing work has explored alongside representation learning to improve retrieval. Context-aware representation learning methods, such as InSeNT and EvoEmbedding, incorporate surrounding information when learning retrieval representations \citep{conti2025context,nie2026evo}. Search-Adaptor instead adapts the final embeddings produced by existing models \citep{yoon2024search}. On the matching side, multi-vector methods such as ColBERT score relevance through token-level matching \citep{khattab2020colbert}. Recent work further shows that existing single-vector models can support such matching by reusing their internal token representations \citep{zhang2026smart}.

\begingroup
\parfillskip=0pt plus 0.3\linewidth
We introduce a memory retrieval method that uses contextual token representations already computed by a frozen embedding model. Specifically, query tokens retrieve sparse matches across candidate memories, and their scores are aggregated by source memory. Combining these scores with whole-memory similarity improves retrieval without additional training. With supervision, a lightweight query adapter further adjusts token-level matching while keeping the encoder and memory index fixed.
\par\endgroup

We evaluate our method across eight retrieval benchmarks and eight embedding backbones spanning multiple model families and sizes. The results show substantial retrieval improvements without additional training. Downstream evaluations further demonstrate higher question-answering accuracy on long-term memory benchmarks. Controlled comparisons and ablations examine sparse matching and whole-memory similarity, while efficiency experiments measure storage and query latency.

Our contributions are threefold:
\begingroup
\settowidth{\leftmargini}{Our}
\addtolength{\leftmargini}{\labelsep}
\begin{itemize}
    \item We show that existing token representations improve memory retrieval without additional training, revealing capabilities underutilized by single-vector scoring.
    \item We combine sparse token matching with whole-memory similarity and provide optional query adaptation that leaves the encoder and memory index unchanged.
    \item We evaluate eight backbones on eight retrieval benchmarks and downstream question answering, examining component contributions and quality--cost trade-offs.
\end{itemize}
\endgroup

\section{Related Work}
\noindent\textbf{Long-term memory and retrieval.}
Long-term memory systems organize past interactions to support question answering. A-Mem links structured notes, while MemoryOS manages memories hierarchically \citep{xu2025amem,kang2025memoryos}. LightMem reduces processing costs through compression, consolidation, and offline updates \citep{fang2026lightmem}. For retrieval, AssoMem combines associative links with relevance and temporal signals \citep{zhang2026assomem}, while EvoEmbedding incorporates evolving context into retrieval representations \citep{nie2026evo}. We instead improve retrieval within a fixed memory bank using token representations from a frozen encoder.

\noindent\textbf{Single-vector representation learning.}
Single-vector models learn a shared embedding space for efficient comparison of queries and candidates. Sentence-BERT uses siamese networks to learn sentence embeddings \citep{reimers2019sbert}, while E5 scales contrastive pre-training with weakly supervised text pairs \citep{wang2022e5}. Qwen3-Embedding builds on pretrained language models to support multilingual, instruction-aware retrieval \citep{zhang2025qwen}. These approaches improve the final embedding through representation learning. Our method instead uses the contextual token states of an existing encoder alongside its final embedding, without changing the encoder.

\noindent\textbf{Contextual and multi-vector retrieval.}
Context-aware chunk methods preserve surrounding information in local representations \citep{gunther2024late,conti2025context}, while BERT-MaxP scores query--passage pairs and takes the maximum \citep{dai2019maxp}. ColBERT and ColBERTv2 use token-level late interaction \citep{khattab2020colbert,santhanam2022colbertv2}. Other methods use compact multi-vector representations \citep{macavaney2025constbert,yu2026mpe,xiao2026metaembed,qin2026compression} or improve late-interaction search \citep{santhanam2022plaid,scheerer2025warp}. SMART reuses frozen hidden states with MaxSim and pooled-score fusion \citep{zhang2026smart}. XTR studies token retrieval and missing-score estimation \citep{lee2023xtr}. Our method applies this approach to frozen token states and combines local scores with whole-memory similarity.

\noindent\textbf{Retrieval adaptation.}
Adaptation can operate on final embeddings or on the token states used for matching. Search-Adaptor learns task-specific mappings on top of existing embeddings \citep{yoon2024search}. Query-only adaptation preserves the document index, as studied in enterprise retrieval \citep{jain2026index}; ERA further aligns query and document embedding spaces before supervised query-side adaptation \citep{maekawa2026era}. SMART explores lightweight adaptation of token representations for late interaction \citep{zhang2026smart}. Our optional low-rank adapter updates only the local query branch, leaving the whole-memory scoring branch and stored memory representations unchanged.

\section{Method}
\begin{figure}[t]
\centering\includegraphics[width=\linewidth]{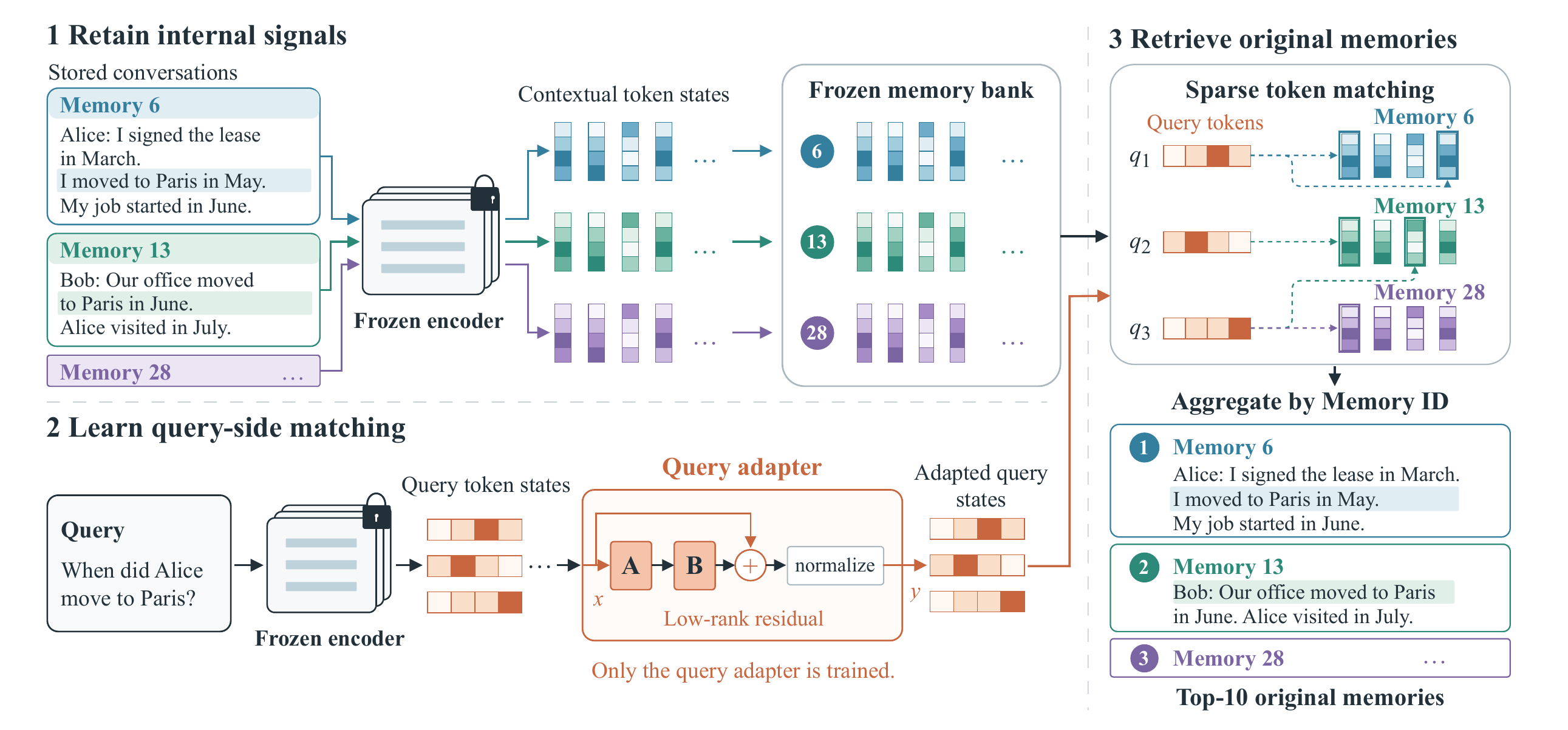}
\caption{\textbf{Sparse token matching for memory retrieval.} Query tokens retrieve matching states from the frozen memory bank; matches are aggregated by source memory and fused with whole-memory similarity for final ranking. The orange query adapter is optional, and memory states and their index remain fixed when it is trained. Memory IDs and alignments are illustrative.}
\label{fig:method}
\end{figure}

\noindent\textbf{Overview.}
Our retrieval method has three stages, as illustrated in Figure~\ref{fig:method}. First, each memory is encoded once by the frozen model, and its contextual token states are stored with their source-memory ID. Second, query states retrieve a sparse set of matching memory tokens, whose scores are aggregated by those IDs. Third, the resulting local evidence is fused with the frozen pooled score to rank the memories. An optional query adapter changes the second stage without modifying the encoder or stored bank.

\subsection{Problem Formulation and the Fixed Memory Bank}
For a query $q$ and its candidate memories $\mathcal{M}_q$, the retriever must rank the original memory records rather than token-level matches. For either input $x\in\{q,m\}$, one forward pass through the frozen encoder provides the backbone's native pooled embedding and its final-layer contextual states. We $\ell_2$-normalize the pooled embedding as $\mathbf{z}_x$ and retain the normalized states at content-token positions as $\mathbf{H}_x$. Thus,
\begin{gather*}
\mathbf{z}_q,\mathbf{z}_m\in\mathbb{R}^d,\\
\mathbf{H}_q=[\mathbf{h}_{q,1},\ldots,\mathbf{h}_{q,n_q}]^{\top}
\in\mathbb{R}^{n_q\times d},\\
\mathbf{H}_m=[\mathbf{h}_{m,1},\ldots,\mathbf{h}_{m,n_m}]^{\top}
\in\mathbb{R}^{n_m\times d}.
\end{gather*}
The original single-vector score is
\begin{gather}
s_{\mathrm{pool}}(q,m)=\mathbf{z}_q^{\top}\mathbf{z}_m.
\label{eq:pool}
\end{gather}
Each memory is encoded independently of the query from its available title and text, so it can be indexed once and reused across queries. We index token states selected by each backbone's content-position mask; Appendix~\ref{app:protocol} records the model-specific prompting, masking, and special-token conventions. Each stored token state is paired with its source-memory ID, allowing token-level matches to be aggregated into rankings over the original memory records.

For a candidate set $\mathcal{M}_q$, the frozen bank contains $N=\sum\nolimits_{m\in\mathcal{M}_q}n_m$ token-state vectors and one pooled embedding per memory. Because the encoder and all memory-side representations remain fixed, the bank can be built once and reused when the query residual is trained or updated. Its representation payload is $O(Nd+|\mathcal{M}_q|d)$, compared with $O(|\mathcal{M}_q|d)$ for a single-vector index. Section~\ref{sec:compression} reports the resulting quality--storage trade-off in measured index bytes.

\subsection{Global Sparse Token Matching}
Every indexed token state $\mathbf{h}_{m,j}$ retains the ID $m$ of its source memory. For each query state $\mathbf{h}_{q,i}$, we use HNSW to approximate global top-$k$ token search across all candidate memories in $\mathcal{M}_q$ and denote the returned set by $\mathcal{R}_i$. Because a returned state remains indexed as $\mathbf{h}_{m,j}$, its memory ID determines which memory receives the corresponding evidence.

Following missing-similarity estimation in token retrieval \citep{lee2023xtr}, we use the clipped similarity of the last retrieved state as the fallback score,
\begin{gather*}
t_i=\left[
\min_{\mathbf{h}\in\mathcal{R}_i}
\mathbf{h}_{q,i}^{\top}\mathbf{h}
\right]_+,
\qquad [u]_+=\max(u,0).
\end{gather*}
For each query token, we take the highest similarity among the retrieved tokens belonging to memory $m$, using $t_i$ as an imputed score floor. Averaging these contributions over query tokens gives the memory score:
\begin{gather}
s_{\mathrm{sparse}}(q,m)=\frac{1}{n_q}\sum\nolimits_i
\max\left\{t_i,
\max\nolimits_{j:\mathbf{h}_{m,j}\in\mathcal{R}_i}
\mathbf{h}_{q,i}^{\top}\mathbf{h}_{m,j}\right\}.
\label{eq:sparse}
\end{gather}
When $\mathcal{R}_i$ contains no state from memory $m$, only the fallback $t_i$ contributes. MaxSim takes the highest token similarity within each memory, whereas sparse matching selects token matches across all candidate memories. For each query token, only retrieved matches above the shared fallback score add to a memory's local score. After token retrieval, our implementation aggregates matches in a dense $n_q\times |U|$ array, where $U$ is the set of matched memory IDs. Table~\ref{tab:k-tail-exact} compares retrieval performance under HNSW and exact top-$k$ search.

\subsection{Local--Global Fusion}
Token-level matching provides local evidence, while the pooled score measures whole-memory similarity. For each branch $b\in\{\mathrm{pool},\mathrm{sparse}\}$, let $\mu_b(q)$ and $\sigma_b(q)$ be the mean and standard deviation of $s_b(q,m)$ over $m\in\mathcal{M}_q$. We standardize each branch as
\begin{gather*}
\widehat{s}_b(q,m)=
\frac{s_b(q,m)-\mu_b(q)}
{\max\left(\sigma_b(q),10^{-6}\right)}.
\end{gather*}
This query-wise standardization places local and pooled scores on comparable scales across backbones and candidate sets. We combine the standardized scores as
\begin{gather}
s_{\mathrm{fused}}(q,m)=\alpha\,\widehat{s}_{\mathrm{pool}}(q,m)
+(1-\alpha)\,\widehat{s}_{\mathrm{sparse}}(q,m),
\label{eq:fusion}
\end{gather}
where $\alpha\in[0,1]$ controls the relative weight of pooled and local scores. The pooled score is computed for every candidate memory, so a memory can appear in the final results even if none of its tokens is retrieved. Combining sparse token matching with the pooled score gives the training-free form of our method. Optional query adaptation modifies only the query token states; the pooled query embedding $\mathbf{z}_q$ remains unchanged.

\subsection{Optional Query Adaptation}
With source supervision, we apply a LoRA-style low-rank residual \citep{hu2022lora} to the encoded query token states, keeping the encoder and memory states fixed. The same bias-free matrices $\mathbf{A}\in\mathbb{R}^{r\times d}$ and $\mathbf{B}\in\mathbb{R}^{d\times r}$ are shared across all query positions. With rank $r=32$, define
\begin{equation}
T_{\theta}(\mathbf{h})
=\frac{\norm{\mathbf{h}}_2
\left(\mathbf{h}+\mathbf{B}\mathbf{A}\mathbf{h}\right)}
{\max\left(\norm{\mathbf{h}+\mathbf{B}\mathbf{A}\mathbf{h}}_2,\epsilon\right)},
\qquad
\widetilde{\mathbf{h}}_{q,i}
=T_{\theta}(\mathbf{h}_{q,i}),
\label{eq:adapt}
\end{equation}
where $\theta=\{\mathbf{A},\mathbf{B}\}$ and $\epsilon$ is a numerical safeguard. There is no intermediate activation. Zero initialization of $\mathbf{B}$ recovers the unmodified query. Outside the numerical safeguard, $T_{\theta}$ preserves the input norm and therefore prevents simple magnitude inflation. The head has $2dr$ parameters. It acts only on the local query branch; neither the pooled query embedding nor any stored memory representation changes.

Since sparse ANN retrieval involves discrete neighbor selection, we train the query head from memory-level labels using MaxSim as a differentiable surrogate. At inference, the adapted query states are used for sparse matching, and the resulting scores are fused with the unchanged pooled scores. The MaxSim training score considers all tokens in each candidate:
\begin{gather}
s_{\mathrm{MaxSim}}(q,m)=\frac{1}{n_q}\sum\nolimits_i
\max\nolimits_j\mathbf{h}_{q,i}^{\top}\mathbf{h}_{m,j}.
\label{eq:maxsim}
\end{gather}
Replacing $\mathbf{h}_{q,i}$ with $\widetilde{\mathbf{h}}_{q,i}$ gives the adapted score $s_{\mathrm{MaxSim}}^{\theta}$. For a source query $q$, let $\mathcal{C}_q$ be its training candidate set and $\mathcal{G}_q\subseteq\mathcal{C}_q$ its positive memories. The head minimizes
\begin{gather}
\mathcal{L}=-\log
\frac{\sum\nolimits_{m\in\mathcal{G}_q}\exp(s_{\mathrm{MaxSim}}^{\theta}(q,m)/\tau)}
{\sum\nolimits_{m\in\mathcal{C}_q}\exp(s_{\mathrm{MaxSim}}^{\theta}(q,m)/\tau)},
\end{gather}
where $\tau$ is the temperature parameter.

\section{Experimental Setup}
\noindent\textbf{Benchmarks.}
We evaluate our method on retrieval and downstream question answering. For retrieval, LoCoMo, LongMemEval, and REALTALK cover conversational memory \citep{maharana2024locomo,wu2025longmemeval,lee2025realtalk}. The other five benchmarks cover financial reports (ESG-Reports), scientific papers (QASPER and PeerQA), biomedical text (CovidQA), and long documents (MLDR) \citep{mace2025vidore,dasigi2021qasper,baumgartner2025peerqa,moller2020covid,chen2024m3}. For downstream question answering, we use LoCoMo and LongMemEval to assess whether retrieval improvements translate into better answers using past conversations.

\noindent\textbf{Baselines.}
Same-backbone baselines include Original embedding, Search-Adaptor \citep{yoon2024search}, Chunk-MaxP, and MaxSim. Chunk-MaxP uses MaxP aggregation \citep{dai2019maxp} over independent chunk representations, taking the highest score per memory. MaxSim applies ColBERT's late interaction to frozen token representations \citep{khattab2020colbert}. BM25 + Original combines BM25 and original embedding scores using our query-wise normalization and equal weighting. Standalone baselines include BM25, ColBERTv2, XTR, and ConstBERT \citep{robertson2009bm25,santhanam2022colbertv2,lee2023xtr,macavaney2025constbert}. Multi-vector systems retain their native encoders and retrieval configurations. \emph{Ours (training-free)} and \emph{Ours} denote our method without and with query adaptation, respectively. The eight frozen backbones are Qwen3-Embedding (0.6B/4B/8B) \citep{zhang2025qwen}, NVIDIA Llama-based models (1B/3B/8B), EmbeddingGemma-300M, and Jina-Embeddings-v5-text-nano. Same-backbone comparisons use identical query and memory inputs.

\noindent\textbf{Implementation Details.}
We average capped Recall@10 and NDCG@10 equally across eight benchmarks. We use uncompressed token indexes and HNSW search \citep{malkov2020hnsw}, with $k=64$ and $\alpha=0.5$. Trainable methods share the EvoTrain-derived source training and validation split \citep{nie2026evo}. Model selection uses only source validation data, and trainable retrieval results average three training seeds. For downstream question answering, we compare Original embedding with our adapted method using Qwen3-30B-A3B-Instruct as the answering model and GPT-4o-mini as the fixed judge. The main comparison retrieves 10 memories; budget analysis varies $K$ over $\{4,8,10,16\}$. Appendices~\ref{app:protocol}, \ref{app:sa}, and~\ref{app:qa} provide detailed protocols.

\section{Results}
\subsection{Main Retrieval Results}
\begin{table}[!t]\centering
\caption{\textbf{Retrieval performance (\%).} Overall is the eight-task mean. Bold and underline denote best and second-best scores within each group; trained methods average three seeds.}\label{tab:native-quality}\label{tab:extension}
\fontsize{8.2}{9.0}\selectfont\setlength{\tabcolsep}{1.1pt}\renewcommand{\arraystretch}{0.92}
\setlength{\aboverulesep}{0.8pt}\setlength{\belowrulesep}{0.8pt}
\newcommand{\resulthead}[2][c]{%
  \makebox[0pt][#1]{\fontsize{7.0}{8.0}\selectfont\rmfamily\bfseries #2}}

\begin{tabularx}{\linewidth}{l>{\hsize=0.8112\hsize\centering\arraybackslash}X>{\hsize=1.3947\hsize\centering\arraybackslash}X>{\hsize=1.0592\hsize\centering\arraybackslash}X>{\hsize=1.2634\hsize\centering\arraybackslash}X>{\hsize=0.7675\hsize\centering\arraybackslash}X>{\hsize=1.1467\hsize\centering\arraybackslash}X>{\hsize=0.7675\hsize\centering\arraybackslash}X>{\hsize=1.0883\hsize\centering\arraybackslash}X>{\hsize=0.7015\hsize\centering\arraybackslash}X}\toprule
\multicolumn{10}{c}{\textbf{(a) Recall@10 $\uparrow$}} \\
\midrule \multicolumn{1}{l}{\resulthead[l]{Method}} & \multicolumn{1}{c}{\resulthead{LoCoMo}} & \multicolumn{1}{c}{\resulthead{LongMemEval}} & \multicolumn{1}{c}{\resulthead{REALTALK}} & \multicolumn{1}{c}{\resulthead{ESG-Reports}} & \multicolumn{1}{c}{\resulthead{QASPER}} & \multicolumn{1}{c}{\resulthead{PeerQA}} & \multicolumn{1}{c}{\resulthead{CovidQA}} & \multicolumn{1}{c}{\resulthead{MLDR}} & \multicolumn{1}{c}{\resulthead{Overall}} \\
\midrule\multicolumn{10}{l}{\textit{Standalone retrieval systems}} \\
BM25 & 48.3 & 81.3 & 40.0 & 50.5 & 61.4 & 30.7 & \underline{94.1} & \textbf{100.0} & 63.3 \\
ColBERTv2 & \underline{61.8} & \underline{86.3} & \underline{52.6} & \underline{56.0} & \textbf{73.4} & \textbf{44.5} & \textbf{94.3} & 97.0 & \textbf{70.8} \\
XTR & 43.6 & \textbf{89.4} & 42.7 & \textbf{62.9} & 71.2 & 38.8 & 93.9 & \underline{98.0} & 67.5 \\
ConstBERT & \textbf{63.1} & 85.7 & \textbf{55.9} & 53.0 & \underline{72.7} & \underline{43.4} & 91.2 & 96.0 & \underline{70.1} \\
\midrule\multicolumn{10}{l}{\textit{Qwen3-Embedding-0.6B}} \\
Original embedding & 46.9 & 84.8 & 46.6 & 52.8 & 69.6 & 36.8 & 92.0 & 95.0 & 65.6 \\
BM25 + Original & 54.3 & 90.5 & 46.5 & 58.7 & 71.4 & 39.0 & \textbf{96.1} & \textbf{98.0} & 69.3 \\
Search-Adaptor & 48.8 & 88.1 & 48.4 & 53.1 & 71.7 & 37.6 & 92.2 & 95.0 & 66.9 \\
Chunk-MaxP & 46.9 & 84.7 & 46.6 & 52.8 & 69.6 & 36.8 & 92.0 & 95.0 & 65.5 \\
MaxSim & 64.2 & 92.2 & 53.4 & 54.3 & 73.8 & 43.0 & 94.4 & \underline{96.0} & 71.4 \\
\rowcolor{black!8}Ours (training-free) & \underline{68.0} & \textbf{94.5} & \underline{55.1} & \underline{64.1} & \underline{76.2} & \textbf{44.6} & 95.4 & \underline{96.0} & \underline{74.2} \\
\rowcolor{black!8}\textbf{Ours} & \textbf{68.7} & \underline{94.1} & \textbf{55.6} & \textbf{65.1} & \textbf{76.6} & \underline{44.3} & \underline{95.9} & \underline{96.0} & \textbf{74.5} \\
\midrule\multicolumn{10}{l}{\textit{Qwen3-Embedding-4B}} \\
Original embedding & 41.6 & 59.2 & 42.8 & 57.9 & 73.1 & 35.8 & 93.3 & \underline{96.0} & 62.5 \\
BM25 + Original & 53.1 & 89.6 & 45.8 & 59.7 & 73.6 & 39.9 & \textbf{96.7} & \textbf{99.0} & 69.7 \\
Search-Adaptor & 45.4 & 84.7 & 45.1 & 57.3 & 74.1 & 38.4 & 94.0 & \underline{96.0} & 66.9 \\
Chunk-MaxP & 41.6 & 59.5 & 42.8 & 57.9 & 73.1 & 35.8 & 93.3 & \underline{96.0} & 62.5 \\
MaxSim & 54.2 & 93.3 & 48.4 & 49.2 & 68.2 & 39.3 & 94.6 & 95.0 & 67.8 \\
\rowcolor{black!8}Ours (training-free) & \underline{63.5} & \underline{95.4} & \underline{52.3} & \underline{68.0} & \underline{75.1} & \textbf{44.2} & 95.7 & \underline{96.0} & \underline{73.8} \\
\rowcolor{black!8}\textbf{Ours} & \textbf{67.8} & \textbf{95.9} & \textbf{56.6} & \textbf{70.8} & \textbf{78.9} & \underline{44.0} & \underline{95.9} & \underline{96.0} & \textbf{75.7} \\
\addlinespace[3pt]
\multicolumn{10}{c}{\textbf{(b) NDCG@10 $\uparrow$}} \\
\midrule \multicolumn{1}{l}{\resulthead[l]{Method}} & \multicolumn{1}{c}{\resulthead{LoCoMo}} & \multicolumn{1}{c}{\resulthead{LongMemEval}} & \multicolumn{1}{c}{\resulthead{REALTALK}} & \multicolumn{1}{c}{\resulthead{ESG-Reports}} & \multicolumn{1}{c}{\resulthead{QASPER}} & \multicolumn{1}{c}{\resulthead{PeerQA}} & \multicolumn{1}{c}{\resulthead{CovidQA}} & \multicolumn{1}{c}{\resulthead{MLDR}} & \multicolumn{1}{c}{\resulthead{Overall}} \\
\midrule\multicolumn{10}{l}{\textit{Standalone retrieval systems}} \\
BM25 & 36.7 & 73.0 & 31.4 & 37.9 & 36.9 & 23.0 & 80.7 & 82.5 & 50.2 \\
ColBERTv2 & \underline{49.4} & \textbf{79.5} & \underline{45.0} & \underline{44.9} & \textbf{51.1} & \textbf{36.0} & \textbf{84.0} & \underline{86.1} & \textbf{59.5} \\
XTR & 31.6 & \underline{79.1} & 33.1 & \textbf{51.9} & 48.8 & 29.8 & \underline{83.3} & \textbf{88.2} & 55.7 \\
ConstBERT & \textbf{50.3} & 76.6 & \textbf{46.7} & 41.5 & \underline{50.1} & \underline{34.7} & 76.8 & 82.6 & \underline{57.4} \\
\midrule\multicolumn{10}{l}{\textit{Qwen3-Embedding-0.6B}} \\
Original embedding & 34.2 & 72.8 & 37.3 & 37.4 & 47.0 & 27.1 & 74.3 & 76.6 & 50.8 \\
BM25 + Original & 42.1 & 81.5 & 38.5 & 44.4 & 46.7 & 29.6 & \underline{84.0} & \textbf{84.5} & 56.4 \\
Search-Adaptor & 35.4 & 76.6 & 38.9 & 38.3 & 49.2 & 27.8 & 75.4 & 76.0 & 52.2 \\
Chunk-MaxP & 34.2 & 72.5 & 37.3 & 37.4 & 47.0 & 27.1 & 74.3 & 76.6 & 50.8 \\
MaxSim & 51.3 & 82.5 & 43.8 & 37.7 & 47.1 & 32.2 & 82.7 & 80.7 & 57.3 \\
\rowcolor{black!8}Ours (training-free) & \underline{53.4} & \underline{87.1} & \underline{44.9} & \textbf{52.3} & \underline{51.7} & \underline{34.0} & 83.8 & 81.6 & \underline{61.1} \\
\rowcolor{black!8}\textbf{Ours} & \textbf{54.5} & \textbf{87.5} & \textbf{45.9} & \underline{51.3} & \textbf{52.5} & \textbf{34.9} & \textbf{84.4} & \underline{81.7} & \textbf{61.6} \\
\midrule\multicolumn{10}{l}{\textit{Qwen3-Embedding-4B}} \\
Original embedding & 29.9 & 47.1 & 34.4 & 40.8 & 50.3 & 27.0 & 76.7 & 77.4 & 48.0 \\
BM25 + Original & 40.6 & 80.6 & 38.3 & 45.5 & 47.9 & 29.8 & 85.0 & \textbf{85.4} & 56.6 \\
Search-Adaptor & 32.9 & 70.9 & 36.4 & 42.6 & \underline{51.4} & 28.5 & 77.8 & 77.7 & 52.3 \\
Chunk-MaxP & 29.9 & 47.4 & 34.4 & 40.8 & 50.2 & 27.0 & 76.7 & 77.4 & 48.0 \\
MaxSim & 43.3 & 84.6 & 40.5 & 31.9 & 42.9 & 30.0 & 82.6 & 79.0 & 54.3 \\
\rowcolor{black!8}Ours (training-free) & \underline{49.7} & \underline{87.7} & \underline{42.7} & \underline{49.3} & 51.1 & \underline{33.7} & \underline{85.3} & 80.0 & \underline{59.9} \\
\rowcolor{black!8}\textbf{Ours} & \textbf{53.6} & \textbf{88.2} & \textbf{46.7} & \textbf{53.3} & \textbf{54.4} & \textbf{34.2} & \textbf{85.5} & \underline{82.5} & \textbf{62.3} \\
\bottomrule\end{tabularx}
\end{table}

\begin{table}[!t]\centering
\caption{\textbf{Cross-backbone retrieval (\%).} Eight-task means; Ours averages three training seeds. Bold and underline mark the best and second-best scores within each backbone and metric.}\label{tab:backbone-summary}
\fontsize{8.2}{9.5}\selectfont\setlength{\tabcolsep}{1.1pt}\renewcommand{\arraystretch}{0.98}
\begin{tabularx}{\linewidth}{l>{\centering\arraybackslash}X*{2}{>{\columncolor{oursblue}\centering\arraybackslash}X}p{3pt}>{\centering\arraybackslash}X*{2}{>{\columncolor{oursblue}\centering\arraybackslash}X}}\toprule
& \multicolumn{3}{c}{\textbf{Recall@10 $\boldsymbol{\uparrow}$}} & & \multicolumn{3}{c}{\textbf{NDCG@10 $\boldsymbol{\uparrow}$}} \\
\cmidrule{2-4}\cmidrule{6-8}
Model & Original & {\fontsize{7.4}{8.6}\selectfont\shortstack{Ours\\(training-free)}} & Ours &  & Original & {\fontsize{7.4}{8.6}\selectfont\shortstack{Ours\\(training-free)}} & Ours \\
\midrule
Qwen3-Embedding-0.6B & 65.6 & \underline{74.2} & \textbf{74.5} &  & 50.8 & \underline{61.1} & \textbf{61.6} \\
Qwen3-Embedding-4B & 62.5 & \underline{73.8} & \textbf{75.7} &  & 48.0 & \underline{59.9} & \textbf{62.3} \\
Llama-Nemotron-Embed-1B-v2 & 74.8 & \underline{77.3} & \textbf{77.6} &  & 61.4 & \underline{65.1} & \textbf{65.4} \\
Llama-NV-Embed-Reasoning-3B & 77.9 & \textbf{79.6} & \underline{79.5} &  & 60.9 & \textbf{67.4} & \underline{66.8} \\
EmbeddingGemma-300M & \underline{72.8} & \textbf{79.8} & \textbf{79.8} &  & \underline{59.4} & \textbf{67.1} & \textbf{67.1} \\
Jina-Embeddings-v5-text-nano & 71.4 & \underline{77.0} & \textbf{77.3} &  & 56.5 & \underline{64.4} & \textbf{64.6} \\
Qwen3-Embedding-8B & 69.1 & \underline{72.9} & \textbf{73.6} &  & 54.1 & \underline{59.1} & \textbf{60.8} \\
Llama-Embed-Nemotron-8B & 76.8 & \underline{79.2} & \textbf{79.3} &  & 62.9 & \underline{67.8} & \textbf{68.1} \\
\bottomrule\end{tabularx}
\end{table}

\noindent\textbf{Overall retrieval performance.}
Table~\ref{tab:native-quality} shows that our method achieves the highest average Recall@10 and NDCG@10 among the evaluated methods with both Qwen3-Embedding-0.6B and Qwen3-Embedding-4B. On the 4B backbone, training-free fusion raises average Recall@10/NDCG@10 from 62.5/48.0 to 73.8/59.9. Query adaptation further raises them to 75.7/62.3. For comparison, direct MaxSim on the same backbone reaches 67.8/54.3 average Recall@10/NDCG@10, while ColBERTv2 achieves 70.8/59.5. Substantial improvements are already obtained without query adaptation, demonstrating the retrieval value of token representations from the frozen encoder.

\noindent\textbf{Gains across conversational and document retrieval.}
Gains over Original embedding are particularly large on conversational memory benchmarks. On Qwen3-Embedding-4B, LoCoMo Recall@10/NDCG@10 rises from 41.6/29.9 to 67.8/53.6, and LongMemEval from 59.2/47.1 to 95.9/88.2. Improvements also extend to document retrieval, including ESG-Reports and QASPER. Compared with standalone systems, our method leads both metrics on LoCoMo and LongMemEval, while ColBERTv2 leads both on PeerQA. On MLDR, BM25 leads Recall@10 and XTR leads NDCG@10.

\noindent\textbf{Retrieval across model families and sizes.}
Table~\ref{tab:backbone-summary} shows that our method improves average Recall@10 and NDCG@10 over Original embedding on all eight backbones. These gains hold with and without query adaptation across Qwen3-Embedding, Llama-based models, EmbeddingGemma, and Jina-Embeddings.

\subsection{Downstream Question Answering}\label{sec:downstream-qa}
Table~\ref{tab:downstream-qa} compares Original embedding and our adapted method across eight backbones. All comparisons use the same answering model, prompts, and top-10 retrieval setting.
\par
\begin{table}[!htbp]\centering
\caption{\textbf{Downstream question answering at top-10 (\%).} Question-type columns report accuracy, while overall columns report accuracy and F1. Bold marks the better score per backbone--Ours pair.}\label{tab:downstream-qa}
\fontsize{7.6}{8.8}\selectfont\setlength{\tabcolsep}{1.1pt}\renewcommand{\arraystretch}{0.98}
\begin{tabularx}{\linewidth}{>{\raggedright\arraybackslash}p{0.27\linewidth}>{\hsize=1.04\hsize\centering\arraybackslash}X>{\hsize=0.97\hsize\centering\arraybackslash}X>{\hsize=1.08\hsize\centering\arraybackslash}X>{\hsize=0.97\hsize\centering\arraybackslash}X>{\hsize=0.97\hsize\centering\arraybackslash}X>{\hsize=0.97\hsize\centering\arraybackslash}X>{\cellcolor{white}}p{4.6pt}*{4}{>{\centering\arraybackslash}X}>{\hsize=1.08\hsize\centering\arraybackslash}X*{2}{>{\centering\arraybackslash}X}>{\hsize=0.92\hsize\centering\arraybackslash}X}\toprule
& \multicolumn{6}{c}{\textbf{LoCoMo}} & & \multicolumn{8}{c}{\textbf{LongMemEval}} \\
\cmidrule{2-7}\cmidrule{9-16}
& \multicolumn{4}{c}{Question type} & \multicolumn{2}{c}{\textbf{Overall}} & & \multicolumn{6}{c}{Question type} & \multicolumn{2}{c}{\textbf{Overall}} \\
Model & S-hop & Multi & Temp. & Open & \textbf{Acc.} & \textbf{F1} &  & User & Asst. & Pref. & Multi & Temp. & Upd. & \textbf{Acc.} & \textbf{F1} \\
\midrule
Qwen3-Embedding-0.6B & 76.1 & 57.1 & 42.7 & 53.1 & 64.2 & 31.2 &  & 85.7 & 94.6 & 56.7 & 45.1 & 50.4 & 82.1 & 64.2 & 12.7 \\
\rowcolor{oursblue}\hspace*{1.2em}\textbf{+ Ours} & \textbf{86.2} & \textbf{60.6} & \textbf{46.7} & \textbf{57.3} & \textbf{71.5} & \textbf{34.1} &  & \textbf{94.3} & \textbf{100.0} & \textbf{60.0} & \textbf{49.6} & \textbf{60.2} & \textbf{85.9} & \textbf{70.6} & \textbf{13.1} \\
Qwen3-Embedding-4B & 76.8 & 55.3 & 38.9 & 52.1 & 63.4 & 31.2 &  & 91.4 & 96.4 & 60.0 & 50.4 & 56.4 & 78.2 & 67.8 & \textbf{13.1} \\
\rowcolor{oursblue}\hspace*{1.2em}\textbf{+ Ours} & \textbf{82.8} & \textbf{55.7} & \textbf{43.3} & \textbf{55.2} & \textbf{67.9} & \textbf{33.0} &  & \textbf{97.1} & \textbf{98.2} & \textbf{70.0} & \textbf{54.9} & \textbf{59.4} & \textbf{85.9} & \textbf{72.6} & 12.4 \\
Llama-Nemotron-Embed-1B-v2 & 78.4 & 50.0 & 42.1 & 47.9 & 63.7 & 31.9 &  & 84.3 & 94.6 & 53.3 & 39.1 & 43.6 & \textbf{83.3} & 60.6 & 12.5 \\
\rowcolor{oursblue}\hspace*{1.2em}\textbf{+ Ours} & \textbf{87.6} & \textbf{61.7} & \textbf{47.4} & \textbf{58.3} & \textbf{72.7} & \textbf{35.8} &  & \textbf{97.1} & \textbf{96.4} & \textbf{70.0} & \textbf{57.1} & \textbf{61.7} & \textbf{83.3} & \textbf{73.2} & \textbf{13.0} \\
Llama-NV-Embed-Reasoning-3B & 35.4 & 27.0 & 18.7 & 36.5 & 30.5 & 14.2 &  & 81.4 & \textbf{98.2} & 70.0 & 37.6 & 48.1 & 78.2 & 61.6 & 11.3 \\
\rowcolor{oursblue}\hspace*{1.2em}\textbf{+ Ours} & \textbf{60.2} & \textbf{34.4} & \textbf{26.8} & \textbf{42.7} & \textbf{47.4} & \textbf{24.0} &  & \textbf{95.7} & 96.4 & \textbf{76.7} & \textbf{50.4} & \textbf{59.4} & \textbf{87.2} & \textbf{71.6} & \textbf{12.8} \\
EmbeddingGemma-300M & 83.8 & 59.9 & 45.8 & \textbf{57.3} & 69.9 & 33.4 &  & 90.0 & 96.4 & \textbf{66.7} & 49.6 & 56.4 & 80.8 & 68.2 & 12.2 \\
\rowcolor{oursblue}\hspace*{1.2em}\textbf{+ Ours} & \textbf{90.2} & \textbf{68.1} & \textbf{46.4} & 56.2 & \textbf{74.9} & \textbf{36.2} &  & \textbf{94.3} & \textbf{98.2} & 63.3 & \textbf{56.4} & \textbf{59.4} & \textbf{82.1} & \textbf{71.6} & \textbf{12.8} \\
Jina-Embeddings-v5-text-nano & 71.2 & 44.3 & 34.9 & 47.9 & 57.3 & 29.0 &  & \textbf{97.1} & \textbf{100.0} & \textbf{73.3} & 52.6 & 53.4 & \textbf{83.3} & 70.4 & 12.7 \\
\rowcolor{oursblue}\hspace*{1.2em}\textbf{+ Ours} & \textbf{87.3} & \textbf{64.9} & \textbf{47.7} & \textbf{56.2} & \textbf{73.0} & \textbf{35.6} &  & \textbf{97.1} & 98.2 & 66.7 & \textbf{53.4} & \textbf{59.4} & \textbf{83.3} & \textbf{71.6} & \textbf{12.8} \\
Qwen3-Embedding-8B & 82.5 & 61.0 & 41.4 & 50.0 & 68.0 & 33.7 &  & 95.7 & 94.6 & \textbf{70.0} & \textbf{63.2} & 59.4 & 84.6 & \textbf{74.0} & 12.9 \\
\rowcolor{oursblue}\hspace*{1.2em}\textbf{+ Ours} & \textbf{87.4} & \textbf{64.2} & \textbf{43.6} & \textbf{56.2} & \textbf{72.1} & \textbf{35.2} &  & \textbf{97.1} & \textbf{98.2} & \textbf{70.0} & 54.1 & \textbf{62.4} & \textbf{89.7} & 73.8 & \textbf{13.1} \\
Llama-Embed-Nemotron-8B & 87.8 & 58.5 & 45.8 & 55.2 & 71.6 & 34.9 &  & \textbf{97.1} & \textbf{100.0} & \textbf{73.3} & 48.1 & 59.4 & 85.9 & 71.2 & 13.1 \\
\rowcolor{oursblue}\hspace*{1.2em}\textbf{+ Ours} & \textbf{89.9} & \textbf{67.7} & \textbf{48.0} & \textbf{58.3} & \textbf{75.1} & \textbf{36.2} &  & 94.3 & 96.4 & \textbf{73.3} & \textbf{66.2} & \textbf{69.2} & \textbf{88.5} & \textbf{78.2} & \textbf{13.2} \\
\bottomrule\end{tabularx}
\end{table}

\par
Across all eight backbones on LoCoMo, our adapted method improves both answer accuracy and F1 over Original embedding, with accuracy gains ranging from 3.51 to 16.95 percentage points. On LongMemEval, overall accuracy improves on seven of eight backbones, with gains of 1.20--12.60 percentage points. On our main Qwen3-Embedding-4B backbone, accuracy increases from 63.44\% to 67.86\% on LoCoMo and from 67.80\% to 72.60\% on LongMemEval.

\subsection{Component Analysis}
We examine component contributions while keeping the encoder and memory bank fixed.
\par
\begin{table}[!htbp]\centering
\caption{\textbf{Component analysis across eight tasks (\%).} Ours combines sparse matching, local--global fusion, and query adaptation. Each variant changes one component. Removing fusion retains only the sparse score. w/o sparse matching uses MaxSim. Adapted rows average three source-training seeds.}\label{tab:mechanism-summary}
\fontsize{8.0}{9.4}\selectfont\setlength{\tabcolsep}{1.1pt}\renewcommand{\arraystretch}{1.08}
\begin{tabularx}{\linewidth}{l*{2}{>{\centering\arraybackslash}X}p{1.5pt}*{2}{>{\centering\arraybackslash}X}}\toprule
 & \multicolumn{2}{c}{Qwen3-Embedding-0.6B} & & \multicolumn{2}{c}{Qwen3-Embedding-4B} \\
\cmidrule{2-3}\cmidrule{5-6}
Method & Recall@10 & NDCG@10 &  & Recall@10 & NDCG@10 \\\midrule
w/o query adaptation & 74.2 & 61.1 &  & 73.8 & 59.9 \\
w/o local--global fusion & 73.5 & 60.1 &  & 74.5 & 60.6 \\
w/o sparse matching & 72.6 & 59.6 &  & 72.4 & 60.1 \\
\rowcolor{oursblue}\textbf{Ours} & \textbf{74.5} & \textbf{61.6} &  & \textbf{75.7} & \textbf{62.3} \\
\bottomrule\end{tabularx}
\end{table}

\par

\noindent\textbf{Local and global information.}
Table~\ref{tab:mechanism-summary} shows that removing local--global fusion reduces mean Recall@10/NDCG@10 by 1.0/1.5 points on Qwen-0.6B and 1.2/1.7 points on Qwen-4B. Together with the gains over Original embedding, this shows that local matching improves retrieval while whole-memory similarity remains complementary.

\noindent\textbf{Sparse token matching.}
Replacing sparse matching with exhaustive MaxSim, keeping query heads and fusion fixed, lowers mean Recall@10/NDCG@10 by 1.9/2.0 points on Qwen-0.6B and 3.4/2.2 points on Qwen-4B. A matched-fusion, single-index comparison without adaptation extends this result to EmbeddingGemma-300M and Llama-1B/3B: sparse matching improves Recall@10 in 18 of 24 model--task combinations and NDCG@10 in 21, with mean gains of 1.29 and 1.48 points, respectively. The NDCG exceptions are Gemma on REALTALK and both Llama sizes on QASPER.

\noindent\textbf{Query adaptation.}
Query adaptation adds 0.3/0.5 points in mean Recall@10/NDCG@10 on Qwen-0.6B and 2.0/2.4 points on Qwen-4B. Its benefit varies across the backbones in Table~\ref{tab:backbone-summary}: Llama-3B shows a small decrease, while EmbeddingGemma is unchanged after rounding.

\noindent\textbf{Missing-score estimation.}
At $k=64$, replacing the tail floor with zero reduces adapted Recall@10 by 1.61--7.14 points and NDCG@10 by 2.57--7.98 points across the two backbones and two datasets. With the floor retained, exact global token search changes Recall@10 by $-0.06$ to $+1.62$ points and improves NDCG@10 by 0.03--1.91 points over HNSW, as shown in Table~\ref{tab:k-tail-exact}. In these settings, estimating unreturned matches matters more than eliminating approximate-search errors.

\subsection{Sensitivity Analysis}
\begin{figure}[!t]
\centering\includegraphics[width=\linewidth]{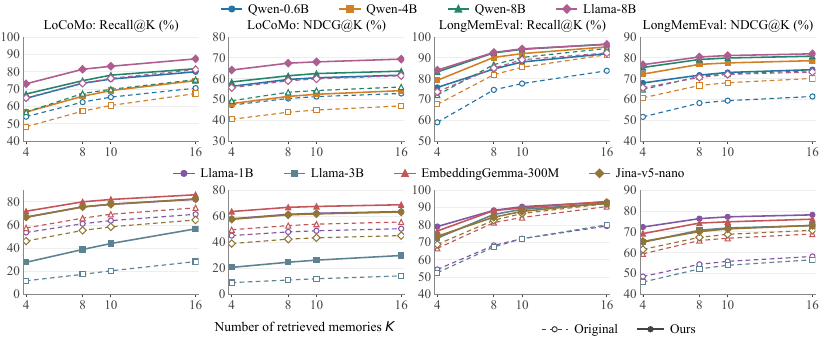}
\plotcaption{\textbf{Native retrieval across memory budgets and eight backbones.} Recall@$K$ and NDCG@$K$ on LoCoMo and LongMemEval for $K\in\{4,8,10,16\}$. Colors identify backbones; dashed lines show Original embedding and solid lines show Ours.}\label{fig:native-retrieval-topk}
\end{figure}

\noindent\textbf{Retrieved memory budget $K$.}
With the retrieval configuration fixed, we vary $K\in\{4,8,10,16\}$ using native memory candidates. Figure~\ref{fig:native-retrieval-topk} shows gains over Original embedding in Recall@$K$ and NDCG@$K$ at every budget across both datasets and all eight backbones. These candidates differ from the main eight-task evaluation and match Table~\ref{tab:downstream-qa} at $K=10$.

\begin{figure}[!t]
\centering\includegraphics[width=\linewidth]{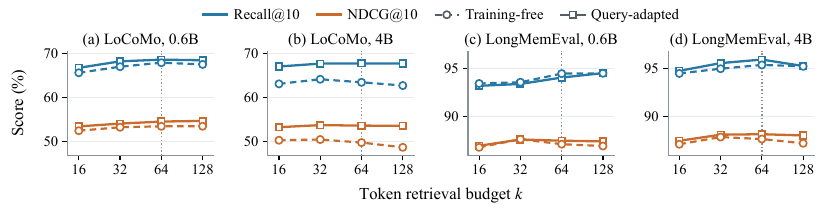}
\plotcaption{\textbf{Sensitivity to the token retrieval budget $k$.} Adapted curves average three training seeds. Dotted lines mark $k=64$.}\label{fig:mechanism}
\end{figure}

\noindent\textbf{Token retrieval budget $k$.}
Figure~\ref{fig:mechanism} shows that retrieving more tokens does not consistently improve Recall@10 or NDCG@10. As $k$ varies from 16 to 128 on LongMemEval, each Recall curve spans less than 1.6 percentage points and each NDCG curve less than 0.9. On LoCoMo, both metrics decline for training-free Qwen-4B when $k$ increases beyond 32.

\begin{figure}[!t]
\centering\includegraphics[width=\linewidth]{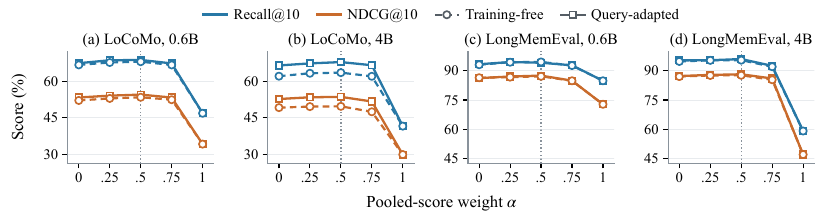}
\plotcaption{\textbf{Sensitivity to the pooled-score weight $\alpha$.} Dotted lines mark $\alpha=0.5$.}\label{fig:fusion}
\end{figure}

\noindent\textbf{Fusion weight.}
Figure~\ref{fig:fusion} shows similar Recall@10/NDCG@10 at $\alpha=0.25$ and $0.5$ across both backbones and datasets, followed by declines at higher weights and a sharp drop at $\alpha=1$.

\subsection{Efficiency and Compression}\label{sec:compression}
Table~\ref{tab:cost-current} compares the uncompressed index (Full), four-bit scalar quantization (SQ4), and 128-dimensional random projection followed by SQ4 (RP128+SQ4) on Qwen-0.6B LoCoMo. Appendix~\ref{app:cost} reports complete compression results, Qwen-4B costs, and the measurement protocol.
\par
\begin{table}[!htbp]\centering
\caption{\textbf{Retrieval quality, storage, and query latency on LoCoMo.}}\label{tab:cost-current}
\fontsize{8.0}{9.4}\selectfont\setlength{\tabcolsep}{1.1pt}\renewcommand{\arraystretch}{1.04}
\begin{tabularx}{\linewidth}{l*{4}{>{\centering\arraybackslash}X}}\toprule
Method & Recall@10 $\uparrow$ & NDCG@10 $\uparrow$ & Size (MiB) $\downarrow$ & Query (ms) $\downarrow$ \\
\midrule
BM25 & 48.28 & 36.68 & 0.23 & 0.76 \\
ColBERTv2 & 61.82 & 49.42 & 1.54 & 27.27 \\
XTR & 43.57 & 31.61 & 5.78 & 93.79 \\
ConstBERT & 63.09 & 50.33 & 4.45 & 10.18 \\
Qwen3-Embedding-0.6B & 46.87 & 34.24 & 2.23 & 16.66 \\
\rowcolor{oursblue}Ours (Full) & 68.53 & 54.45 & 153.93 & 29.21 \\
\rowcolor{oursblue}Ours (SQ4) & 68.37 & 54.31 & 26.17 & 21.68 \\
\rowcolor{oursblue}Ours (RP128+SQ4) & 66.69 & 51.92 & 10.70 & 20.69 \\
\bottomrule\end{tabularx}
\end{table}

\par

Across two models and eight tasks, SQ4 reduces storage by \CostReductionLow--\CostReductionHigh\% relative to Full, with mean Recall@10/NDCG@10 changing by \SQFourRecallMeanDelta{}/\SQFourMeanDelta{} points. On Qwen-0.6B LoCoMo, it retains gains over Original embedding at about 5 ms additional latency. RP128+SQ4 further reduces storage at a larger quality cost.

\section{Conclusion}
Existing contextual token representations support local matching, but single-vector scoring does not directly use them. Therefore, we introduced a method that combines sparse token matching with whole-memory similarity using a frozen embedding model. Experiments across eight models and eight benchmarks showed gains in Recall@10 and NDCG@10 without additional training. Optional query adaptation kept the encoder and memory index fixed, and the adapted method also improved downstream question-answering accuracy. Compression reduced storage and latency relative to the uncompressed index while retaining most retrieval gains. Our method thus made fuller use of the fine-grained retrieval capabilities of existing embedding models.

\bibliography{references}

\begin{thebibliography}{43}
\providecommand{\natexlab}[1]{#1}
\providecommand{\url}[1]{\texttt{#1}}
\expandafter\ifx\csname urlstyle\endcsname\relax
  \providecommand{\doi}[1]{doi: #1}\else
  \providecommand{\doi}{doi: \begingroup \urlstyle{rm}\Url}\fi

\bibitem[Akram et~al.(2026)Akram, Sturua, Havriushenko, Herreros, G{\"u}nther,
  Werk, and Xiao]{akram2026jina}
Mohammad~Kalim Akram, Saba Sturua, Nastia Havriushenko, Quentin Herreros,
  Michael G{\"u}nther, Maximilian Werk, and Han Xiao.
\newblock {jina-embeddings-v5-text}: Task-targeted embedding distillation.
\newblock \emph{arXiv preprint arXiv:2602.15547}, 2026.

\bibitem[Babakhin et~al.(2025)Babakhin, Osmulski, Ak, Moreira, Xu, Schifferer,
  Liu, and Oldridge]{babakhin2025nemotron}
Yauhen Babakhin, Radek Osmulski, Ronay Ak, Gabriel Moreira, Mengyao Xu,
  Benedikt Schifferer, Bo~Liu, and Even Oldridge.
\newblock {Llama-Embed-Nemotron-8B}: A universal text embedding model for
  multilingual and cross-lingual tasks.
\newblock \emph{arXiv preprint arXiv:2511.07025}, 2025.

\bibitem[Baumg{\"a}rtner et~al.(2025)Baumg{\"a}rtner, Briscoe, and
  Gurevych]{baumgartner2025peerqa}
Tim Baumg{\"a}rtner, Ted Briscoe, and Iryna Gurevych.
\newblock {PeerQA}: A scientific question answering dataset from peer reviews.
\newblock In \emph{Proceedings of the 2025 Conference of the Nations of the
  Americas Chapter of the Association for Computational Linguistics: Human
  Language Technologies (Volume 1: Long Papers)}, 2025.

\bibitem[Chen et~al.(2024)Chen, Xiao, Zhang, Luo, Lian, and Liu]{chen2024m3}
Jianlv Chen, Shitao Xiao, Peitian Zhang, Kun Luo, Defu Lian, and Zheng Liu.
\newblock {M3}-embedding: Multi-linguality, multi-functionality,
  multi-granularity text embeddings through self-knowledge distillation.
\newblock In \emph{Findings of the Association for Computational Linguistics:
  ACL 2024}, 2024.

\bibitem[Conti et~al.(2025)Conti, Faysse, Viaud, Bosselut, Hudelot, and
  Colombo]{conti2025context}
Max Conti, Manuel Faysse, Gautier Viaud, Antoine Bosselut, C{\'e}line Hudelot,
  and Pierre Colombo.
\newblock Context is gold to find the gold passage: Evaluating and training
  contextual document embeddings.
\newblock In \emph{Proceedings of the 2025 Conference on Empirical Methods in
  Natural Language Processing}, 2025.

\bibitem[Dai \& Callan(2019)Dai and Callan]{dai2019maxp}
Zhuyun Dai and Jamie Callan.
\newblock Deeper text understanding for {IR} with contextual neural language
  modeling.
\newblock In \emph{Proceedings of the 42nd International ACM SIGIR Conference
  on Research and Development in Information Retrieval}, 2019.

\bibitem[Dasigi et~al.(2021)Dasigi, Lo, Beltagy, Cohan, Smith, and
  Gardner]{dasigi2021qasper}
Pradeep Dasigi, Kyle Lo, Iz~Beltagy, Arman Cohan, Noah~A. Smith, and Matt
  Gardner.
\newblock A dataset of information-seeking questions and answers anchored in
  research papers.
\newblock In \emph{Proceedings of the 2021 Conference of the North American
  Chapter of the Association for Computational Linguistics: Human Language
  Technologies}, 2021.

\bibitem[Esfandiarpoor et~al.(2026)Esfandiarpoor, Osmulski, Babakhin, de~Souza
  P.~Moreira, Holworthy, He, Ak, Cai, Chesler, Liu, and
  Oldridge]{esfandiarpoor2026semanticsimilarityperformancecosts}
Reza Esfandiarpoor, Radek Osmulski, Yauhen Babakhin, Gabriel de~Souza
  P.~Moreira, Oliver Holworthy, Jie He, Ronay Ak, Jiarui Cai, Ryan Chesler,
  Bo~Liu, and Even Oldridge.
\newblock Beyond semantic similarity: Performance and costs of agentic
  retrieval for complex tasks.
\newblock \emph{arXiv preprint arXiv:2610.05750}, 2026.

\bibitem[Fang et~al.(2026)Fang, Deng, Xu, Jiang, Tang, Xu, Deng, Yao, Wang,
  Qiao, Chen, and Zhang]{fang2026lightmem}
Jizhan Fang, Xinle Deng, Haoming Xu, Ziyan Jiang, Yuqi Tang, Ziwen Xu, Shumin
  Deng, Yunzhi Yao, Mengru Wang, Shuofei Qiao, Huajun Chen, and Ningyu Zhang.
\newblock {LightMem}: Lightweight and efficient memory-augmented generation.
\newblock In \emph{International Conference on Learning Representations}, 2026.

\bibitem[G{\"u}nther et~al.(2024)G{\"u}nther, Mohr, Williams, Wang, and
  Xiao]{gunther2024late}
Michael G{\"u}nther, Isabelle Mohr, Daniel~James Williams, Bo~Wang, and Han
  Xiao.
\newblock Late chunking: Contextual chunk embeddings using long-context
  embedding models.
\newblock \emph{arXiv preprint arXiv:2409.04701}, 2024.

\bibitem[Hu et~al.(2022)Hu, Shen, Wallis, Allen-Zhu, Li, Wang, Wang, and
  Chen]{hu2022lora}
Edward~J. Hu, Yelong Shen, Phillip Wallis, Zeyuan Allen-Zhu, Yuanzhi Li, Shean
  Wang, Lu~Wang, and Weizhu Chen.
\newblock {LoRA}: Low-rank adaptation of large language models.
\newblock In \emph{International Conference on Learning Representations}, 2022.

\bibitem[Jain et~al.(2026)Jain, Nair, Goru, Agarwal, Yadav, Varadharajan, and
  Caramanis]{jain2026index}
Prateek Jain, Shabari~S Nair, Ritesh Goru, Prakhar Agarwal, Ajay Yadav, Yoga
  Sri~Varshan Varadharajan, and Constantine Caramanis.
\newblock Succeeding at scale: Enterprise retrieval benchmark construction and
  index-preserving query adaptation for multi-tenant search.
\newblock \emph{arXiv preprint arXiv:2601.04646}, 2026.

\bibitem[Kang et~al.(2025)Kang, Ji, Zhao, and Bai]{kang2025memoryos}
Jiazheng Kang, Mingming Ji, Zhe Zhao, and Ting Bai.
\newblock Memory {OS} of {AI} agent.
\newblock In \emph{Proceedings of the 2025 Conference on Empirical Methods in
  Natural Language Processing}, 2025.

\bibitem[Karpukhin et~al.(2020)Karpukhin, Oguz, Min, Lewis, Wu, Edunov, Chen,
  and tau Yih]{karpukhin2020dpr}
Vladimir Karpukhin, Barlas Oguz, Sewon Min, Patrick Lewis, Ledell Wu, Sergey
  Edunov, Danqi Chen, and Wen tau Yih.
\newblock Dense passage retrieval for open-domain question answering.
\newblock In \emph{Proceedings of the 2020 Conference on Empirical Methods in
  Natural Language Processing (EMNLP)}, 2020.

\bibitem[Khattab \& Zaharia(2020)Khattab and Zaharia]{khattab2020colbert}
Omar Khattab and Matei Zaharia.
\newblock {ColBERT}: Efficient and effective passage search via contextualized
  late interaction over {BERT}.
\newblock In \emph{Proceedings of the 43rd International ACM SIGIR Conference
  on Research and Development in Information Retrieval}, 2020.

\bibitem[Lee et~al.(2025)Lee, Maharana, Pujara, Ren, and
  Barbieri]{lee2025realtalk}
Dong-Ho Lee, Adyasha Maharana, Jay Pujara, Xiang Ren, and Francesco Barbieri.
\newblock {REALTALK}: A 21-day real-world dataset for long-term conversation.
\newblock \emph{arXiv preprint arXiv:2502.13270}, 2025.

\bibitem[Lee et~al.(2023)Lee, Dai, Duddu, Lei, Naim, Chang, and
  Zhao]{lee2023xtr}
Jinhyuk Lee, Zhuyun Dai, Sai Meher~Karthik Duddu, Tao Lei, Iftekhar Naim,
  Ming-Wei Chang, and Vincent~Y. Zhao.
\newblock Rethinking the role of token retrieval in multi-vector retrieval.
\newblock In \emph{Advances in Neural Information Processing Systems},
  volume~36, 2023.

\bibitem[MacAvaney et~al.(2025)MacAvaney, Mallia, and
  Tonellotto]{macavaney2025constbert}
Sean MacAvaney, Antonio Mallia, and Nicola Tonellotto.
\newblock Efficient constant-space multi-vector retrieval.
\newblock In \emph{Advances in Information Retrieval}, 2025.

\bibitem[Mac{\'e} et~al.(2025)Mac{\'e}, Loison, and Faysse]{mace2025vidore}
Quentin Mac{\'e}, Ant{\'o}nio Loison, and Manuel Faysse.
\newblock {ViDoRe} benchmark {V2}: Raising the bar for visual retrieval.
\newblock \emph{arXiv preprint arXiv:2505.17166}, 2025.

\bibitem[Maekawa et~al.(2026)Maekawa, Aminnaseri, Pezeshkpour, and
  Hruschka]{maekawa2026era}
Seiji Maekawa, Moin Aminnaseri, Pouya Pezeshkpour, and Estevam Hruschka.
\newblock Align then adapt: Label-efficient adapter learning for asymmetric
  dense retrieval.
\newblock \emph{arXiv preprint arXiv:2604.03403}, 2026.

\bibitem[Maharana et~al.(2024)Maharana, Lee, Tulyakov, Bansal, Barbieri, and
  Fang]{maharana2024locomo}
Adyasha Maharana, Dong-Ho Lee, Sergey Tulyakov, Mohit Bansal, Francesco
  Barbieri, and Yuwei Fang.
\newblock Evaluating very long-term conversational memory of {LLM} agents.
\newblock In \emph{Proceedings of the 62nd Annual Meeting of the Association
  for Computational Linguistics (Volume 1: Long Papers)}, 2024.

\bibitem[Malkov \& Yashunin(2020)Malkov and Yashunin]{malkov2020hnsw}
Yu.~A. Malkov and D.~A. Yashunin.
\newblock Efficient and robust approximate nearest neighbor search using
  hierarchical navigable small world graphs.
\newblock \emph{IEEE Transactions on Pattern Analysis and Machine
  Intelligence}, 42\penalty0 (4):\penalty0 824--836, 2020.

\bibitem[M{\"o}ller et~al.(2020)M{\"o}ller, Reina, Jayakumar, and
  Pietsch]{moller2020covid}
Timo M{\"o}ller, Anthony Reina, Raghavan Jayakumar, and Malte Pietsch.
\newblock {COVID-QA}: A question answering dataset for {COVID}-19.
\newblock In \emph{Proceedings of the 1st Workshop on {NLP} for {COVID-19} at
  {ACL} 2020}, 2020.

\bibitem[Moreira et~al.(2024)Moreira, Osmulski, Xu, Ak, Schifferer, and
  Oldridge]{moreira2024nvretriever}
Gabriel de Souza~P. Moreira, Radek Osmulski, Mengyao Xu, Ronay Ak, Benedikt
  Schifferer, and Even Oldridge.
\newblock {NV-Retriever}: Improving text embedding models with effective
  hard-negative mining.
\newblock \emph{arXiv preprint arXiv:2407.15831}, 2024.

\bibitem[Nie et~al.(2026)Nie, Fu, Feng, and Shan]{nie2026evo}
Chang Nie, Chaoyou Fu, Junlan Feng, and Caifeng Shan.
\newblock {EvoEmbedding}: Evolvable representations for long-context retrieval
  and agentic memory.
\newblock \emph{arXiv preprint arXiv:2606.21649}, 2026.

\bibitem[Qin et~al.(2026)Qin, Martin, Jha, Zuo, Kriz, and
  Durme]{qin2026compression}
Hanxiang Qin, Alexander Martin, Rohan Jha, Chunsheng Zuo, Reno Kriz, and
  Benjamin~Van Durme.
\newblock Multi-vector index compression in any modality.
\newblock \emph{arXiv preprint arXiv:2602.21202}, 2026.

\bibitem[Reimers \& Gurevych(2019)Reimers and Gurevych]{reimers2019sbert}
Nils Reimers and Iryna Gurevych.
\newblock Sentence-{BERT}: Sentence embeddings using siamese {BERT}-networks.
\newblock In \emph{Proceedings of the 2019 Conference on Empirical Methods in
  Natural Language Processing and the 9th International Joint Conference on
  Natural Language Processing (EMNLP-IJCNLP)}, 2019.

\bibitem[Robertson \& Zaragoza(2009)Robertson and Zaragoza]{robertson2009bm25}
Stephen Robertson and Hugo Zaragoza.
\newblock The probabilistic relevance framework: {BM25} and beyond.
\newblock \emph{Foundations and Trends in Information Retrieval}, 3\penalty0
  (4):\penalty0 333--389, 2009.

\bibitem[Santhanam et~al.(2022{\natexlab{a}})Santhanam, Khattab, Potts, and
  Zaharia]{santhanam2022plaid}
Keshav Santhanam, Omar Khattab, Christopher Potts, and Matei Zaharia.
\newblock {PLAID}: An efficient engine for late interaction retrieval.
\newblock In \emph{Proceedings of the 31st ACM International Conference on
  Information \& Knowledge Management}, 2022{\natexlab{a}}.

\bibitem[Santhanam et~al.(2022{\natexlab{b}})Santhanam, Khattab, Saad-Falcon,
  Potts, and Zaharia]{santhanam2022colbertv2}
Keshav Santhanam, Omar Khattab, Jon Saad-Falcon, Christopher Potts, and Matei
  Zaharia.
\newblock {ColBERTv2}: Effective and efficient retrieval via lightweight late
  interaction.
\newblock In \emph{Proceedings of the 2022 Conference of the North American
  Chapter of the Association for Computational Linguistics: Human Language
  Technologies}, 2022{\natexlab{b}}.

\bibitem[Schechter~Vera et~al.(2025)Schechter~Vera, Dua, Zhang, Salz, Mullins,
  Panyam, Smoot, Naim, Zou, Chen, et~al.]{schechtervera2025embeddinggemma}
Henrique Schechter~Vera, Sahil Dua, Biao Zhang, Daniel Salz, Ryan Mullins,
  Sindhu~Raghuram Panyam, Sara Smoot, Iftekhar Naim, Joe Zou, Feiyang Chen,
  et~al.
\newblock {EmbeddingGemma}: Powerful and lightweight text representations.
\newblock \emph{arXiv preprint arXiv:2509.20354}, 2025.

\bibitem[Scheerer et~al.(2025)Scheerer, Zaharia, Potts, Alonso, and
  Khattab]{scheerer2025warp}
Jan~Luca Scheerer, Matei Zaharia, Christopher Potts, Gustavo Alonso, and Omar
  Khattab.
\newblock {WARP}: An efficient engine for multi-vector retrieval.
\newblock In \emph{Proceedings of the 48th International ACM SIGIR Conference
  on Research and Development in Information Retrieval}, 2025.

\bibitem[Wang et~al.(2022)Wang, Yang, Huang, Jiao, Yang, Jiang, Majumder, and
  Wei]{wang2022e5}
Liang Wang, Nan Yang, Xiaolong Huang, Binxing Jiao, Linjun Yang, Daxin Jiang,
  Rangan Majumder, and Furu Wei.
\newblock Text embeddings by weakly-supervised contrastive pre-training.
\newblock \emph{arXiv preprint arXiv:2212.03533}, 2022.

\bibitem[Weller et~al.(2026)Weller, Boratko, Naim, and
  Lee]{weller2026limitations}
Orion Weller, Michael Boratko, Iftekhar Naim, and Jinhyuk Lee.
\newblock On the theoretical limitations of embedding-based retrieval.
\newblock In \emph{International Conference on Learning Representations}, 2026.

\bibitem[Wu et~al.(2025)Wu, Wang, Yu, Zhang, Chang, and Yu]{wu2025longmemeval}
Di~Wu, Hongwei Wang, Wenhao Yu, Yuwei Zhang, Kai-Wei Chang, and Dong Yu.
\newblock {LongMemEval}: Benchmarking chat assistants on long-term interactive
  memory.
\newblock In \emph{International Conference on Learning Representations}, 2025.

\bibitem[Xiao et~al.(2026)Xiao, Ma, Gu, Chen, Chen, Ordonez, and
  Mohan]{xiao2026metaembed}
Zilin Xiao, Qi~Ma, Mengting Gu, Chun-cheng~Jason Chen, Xintao Chen, Vicente
  Ordonez, and Vijai Mohan.
\newblock {MetaEmbed}: Scaling multimodal retrieval at test-time with flexible
  late interaction.
\newblock In \emph{International Conference on Learning Representations}, 2026.

\bibitem[Xu et~al.(2025)Xu, Liang, Mei, Gao, Tan, and Zhang]{xu2025amem}
Wujiang Xu, Zujie Liang, Kai Mei, Hang Gao, Juntao Tan, and Yongfeng Zhang.
\newblock {A-Mem}: Agentic memory for {LLM} agents.
\newblock In \emph{Advances in Neural Information Processing Systems},
  volume~38, 2025.

\bibitem[Yoon et~al.(2024)Yoon, Chen, Arik, and Pfister]{yoon2024search}
Jinsung Yoon, Yanfei Chen, Sercan~{\"O}. Arik, and Tomas Pfister.
\newblock {Search-Adaptor}: Embedding customization for information retrieval.
\newblock In \emph{Proceedings of the 62nd Annual Meeting of the Association
  for Computational Linguistics (Volume 1: Long Papers)}, 2024.

\bibitem[Yu et~al.(2026)Yu, Ma, Zhuang, Xu, Gao, Zhang, and Lin]{yu2026mpe}
Zhenglin Yu, Xueguang Ma, Shengyao Zhuang, Zhichao Xu, Luyu Gao, Crystina
  Zhang, and Jimmy Lin.
\newblock Improving long-context retrieval with multi-prefix embedding.
\newblock \emph{arXiv preprint arXiv:2606.23642}, 2026.

\bibitem[Zhang et~al.(2026{\natexlab{a}})Zhang, Lee, Ganguly, Kam, Kim, and
  Lee]{zhang2026smart}
Jianrui Zhang, Hyun~Jung Lee, Sukanta Ganguly, Tae-Eui Kam, Donghyun Kim, and
  Yong~Jae Lee.
\newblock Your embedding model is {SMART}er than you think.
\newblock \emph{arXiv preprint arXiv:2605.24938}, 2026{\natexlab{a}}.

\bibitem[Zhang et~al.(2026{\natexlab{b}})Zhang, Zhang, Ahmed, Jiang, Kumar,
  Sun, Lin, Sharma, Oraby, Colak, Aly, Kumar, Liu, and Dong]{zhang2026assomem}
Kai Zhang, Xinyuan Zhang, Ejaz Ahmed, Hongda Jiang, Caleb Kumar, Kai Sun,
  Zhaojiang Lin, Sanat Sharma, Shereen Oraby, Aaron Colak, Ahmed~A. Aly, Anuj
  Kumar, Xiaozhong Liu, and Xin~Luna Dong.
\newblock {AssoMem}: Scalable memory {QA} with multi-signal associative
  retrieval.
\newblock In \emph{International Conference on Learning Representations},
  2026{\natexlab{b}}.

\bibitem[Zhang et~al.(2025)Zhang, Li, Long, Zhang, Lin, Yang, Xie, Yang, Liu,
  Lin, Huang, and Zhou]{zhang2025qwen}
Yanzhao Zhang, Mingxin Li, Dingkun Long, Xin Zhang, Huan Lin, Baosong Yang,
  Pengjun Xie, An~Yang, Dayiheng Liu, Junyang Lin, Fei Huang, and Jingren Zhou.
\newblock {Qwen3} embedding: Advancing text embedding and reranking through
  foundation models.
\newblock \emph{arXiv preprint arXiv:2506.05176}, 2025.

\bibitem[Zhao et~al.(2026)Zhao, Hu, Xu, Tang, Zhang, Zhou, Zhong, Zhou, Shan,
  Zhang, Hu, and Zhang]{zhao2026lmeb}
Xinping Zhao, Xinshuo Hu, Jiaxin Xu, Danyu Tang, Xin Zhang, Mengjia Zhou, Yan
  Zhong, Yao Zhou, Zifei Shan, Meishan Zhang, Baotian Hu, and Min Zhang.
\newblock {LMEB}: Long-horizon memory embedding benchmark.
\newblock \emph{arXiv preprint arXiv:2603.12572}, 2026.

\end{thebibliography}
\bibliographystyle{preprint}
\newpage
\appendix
\section{Evaluation Details}\label{app:protocol}
We follow the LMEB evaluation protocol \citep{zhao2026lmeb} on 5,873 queries across eight benchmarks and report capped Recall@10 and NDCG@10 as percentages. We average subtasks equally within each benchmark and report the eight-benchmark mean. Retrieval metrics are computed from each system's returned rankings. Table~\ref{tab:encoder-protocol} lists encoder settings.
\begin{table}[ht]
\centering
\caption{\textbf{Backbone encoding settings.}}
\label{tab:encoder-protocol}
\fontsize{8.2}{10}\selectfont
\setlength{\tabcolsep}{1.1pt}
\renewcommand{\arraystretch}{1.08}
\begin{tabular*}{\linewidth}{@{\extracolsep{\fill}\hspace{\tabcolsep}}lllr@{\hspace{\tabcolsep}}}
\toprule
Model & Dimension & Whole-vector pooling & Input limit \\
\midrule
Qwen-0.6B / 4B & 1024 / 2560 & Last valid token & 32,768 \\
Qwen-8B & 4096 & Last valid token & 32,768 \\
Llama-1B / 3B & 2048 / 3072 & Masked mean & 8,192 \\
Llama-8B & 4096 & Masked mean & 32,768 \\
EmbeddingGemma & 768 & Masked mean + projection & 2,048 \\
Jina-v5-text-nano & 768 & Last valid token & 8,192 \\
\bottomrule
\end{tabular*}
\end{table}

\noindent\textbf{Model sources.}
Llama-1B, Llama-3B, and Llama-8B refer to \href{https://huggingface.co/nvidia/llama-nemotron-embed-1b-v2}{Llama-Nemotron-Embed-1B-v2} \citep{moreira2024nvretriever}, \href{https://huggingface.co/nvidia/llama-nv-embed-reasoning-3b}{Llama-NV-Embed-Reasoning-3B} \citep{esfandiarpoor2026semanticsimilarityperformancecosts}, and \href{https://huggingface.co/nvidia/llama-embed-nemotron-8b}{Llama-Embed-Nemotron-8B} \citep{babakhin2025nemotron}, respectively. We also use \href{https://huggingface.co/google/embeddinggemma-300m}{EmbeddingGemma-300M} \citep{schechtervera2025embeddinggemma} and \href{https://huggingface.co/jinaai/jina-embeddings-v5-text-nano}{Jina-Embeddings-v5-text-nano} \citep{akram2026jina}. Model names link to their official model cards.

\noindent\textbf{Encoding and token search.}
We follow each backbone's native encoding formats and implementations, adding LMEB task instructions for Qwen and Llama-8B. Local matching uses content-token states, excluding prefixes, padding, and special tokens; whole-memory vectors retain native pooling. EmbeddingGemma token states are taken before its pooled projection. Encoders use BF16 except EmbeddingGemma (FP32), and normalized token states are stored in FP16. Token retrieval uses an inner-product HNSW index with $M=16$, \texttt{efConstruction}=80, \texttt{efSearch}=128, and one search thread.

\section{Training Details}\label{app:sa}
Our query adapter and Search-Adaptor use 1,024 training and 256 validation examples from EvoTrain-180K \citep{nie2026evo}. Our adapter uses the settings in Table~\ref{tab:adapter-settings} across backbones, with all candidates in each query's source bank contributing to the training loss. We select its checkpoint by the mean of source-validation Recall@10 and NDCG@10 at epochs 0, 1, 5, 10, 15, 20, and 25. Trainable retrieval results average seeds 31070, 31071, and 31072.
\begin{table}[!htbp]\centering
\caption{\textbf{Query-adapter training settings.}}\label{tab:adapter-settings}
\fontsize{8.2}{10}\selectfont\setlength{\tabcolsep}{4pt}\renewcommand{\arraystretch}{1.06}
\begin{tabularx}{\linewidth}{@{\hspace{1.1pt}}lX@{\hspace{1.1pt}}}\toprule
Setting & Value \\
\midrule
Adapter & Rank-32 query residual; zero-initialized output matrix \\
Optimizer & AdamW; learning rate $10^{-4}$; weight decay $10^{-4}$ \\
Training budget & 25 epochs; batch size 16 queries \\
Loss & Multi-positive MaxSim cross-entropy; temperature $\tau=0.07$ \\
Gradient clipping & Maximum norm 1 \\
\bottomrule\end{tabularx}
\end{table}

\section{Per-Task Backbone Results}\label{app:tasks}
\begin{table}[H]\centering
\caption{\textbf{Per-task capped Recall@10 for the six additional backbones in Table~\ref{tab:backbone-summary} (\%).} Backbone rows report Original embedding; Ours reports mean $\pm$ SD.}
\fontsize{8.0}{9.0}\selectfont\setlength{\tabcolsep}{1.1pt}\renewcommand{\arraystretch}{0.96}
\begin{tabularx}{\linewidth}{l*{7}{>{\centering\arraybackslash}X}>{\centering\arraybackslash}X}\toprule
Model & LC & LM & RT & ESG & QA & PQ & CQ & MLDR \\
\midrule\textbf{Llama-1B} & 68.5 & 93.6 & 57.2 & 60.6 & 77.5 & 46.6 & 95.5 & \textbf{99.0} \\
\rowcolor{oursblue}Ours (training-free) & 72.3 & \textbf{95.7} & 60.5 & 67.1 & 79.9 & 47.7 & 97.4 & 98.0 \\
\rowcolor{oursblue}\textbf{Ours} & \textbf{72.6}\smallstd{0.14} & 95.5\smallstd{0.11} & \textbf{61.2}\smallstd{0.12} & \textbf{68.0}\smallstd{0.80} & \textbf{80.2}\smallstd{0.14} & \textbf{47.8}\smallstd{0.47} & \textbf{97.5}\smallstd{0.05} & 98.0\smallstd{0.00} \\
\midrule\textbf{Llama-3B} & 72.4 & 89.4 & 61.2 & \textbf{81.3} & 80.7 & 45.8 & 93.8 & \textbf{99.0} \\
\rowcolor{oursblue}Ours (training-free) & 76.2 & \textbf{95.7} & \textbf{64.9} & 75.6 & 81.9 & \textbf{47.4} & 97.1 & 98.0 \\
\rowcolor{oursblue}\textbf{Ours} & \textbf{76.6}\smallstd{0.17} & 94.8\smallstd{0.42} & 63.9\smallstd{0.23} & 75.8\smallstd{1.06} & \textbf{82.2}\smallstd{0.18} & 47.2\smallstd{0.47} & \textbf{97.2}\smallstd{0.05} & 98.0\smallstd{0.00} \\
\midrule\textbf{EmbeddingGemma-300M} & 64.8 & 91.9 & 58.5 & 50.5 & 78.6 & 45.5 & 93.6 & \textbf{99.0} \\
\rowcolor{oursblue}Ours (training-free) & 76.1 & \textbf{95.7} & 61.4 & \textbf{78.3} & 80.9 & 50.1 & \textbf{97.0} & \textbf{99.0} \\
\rowcolor{oursblue}\textbf{Ours} & \textbf{76.4}\smallstd{0.11} & \textbf{95.7}\smallstd{0.07} & \textbf{61.9}\smallstd{0.07} & 76.9\smallstd{0.80} & \textbf{81.0}\smallstd{0.04} & \textbf{50.2}\smallstd{0.00} & \textbf{97.0}\smallstd{0.00} & \textbf{99.0}\smallstd{0.00} \\
\midrule\textbf{Jina-v5-text-nano} & 58.4 & 88.7 & 52.5 & 65.9 & 71.4 & 42.0 & 94.3 & \textbf{98.0} \\
\rowcolor{oursblue}Ours (training-free) & 71.7 & \textbf{95.0} & 58.8 & \textbf{74.3} & 78.2 & 44.0 & 97.0 & 97.0 \\
\rowcolor{oursblue}\textbf{Ours} & \textbf{72.3}\smallstd{0.06} & \textbf{95.0}\smallstd{0.11} & \textbf{59.3}\smallstd{0.23} & \textbf{74.3}\smallstd{0.00} & \textbf{78.4}\smallstd{0.30} & \textbf{45.2}\smallstd{1.31} & \textbf{97.1}\smallstd{0.05} & 97.0\smallstd{0.00} \\
\midrule\textbf{Qwen3-Embedding-8B} & 49.5 & 87.9 & 50.5 & 65.0 & 70.4 & 40.7 & 93.9 & 95.0 \\
\rowcolor{oursblue}Ours (training-free) & 60.6 & 94.1 & 51.5 & \textbf{65.1} & 75.5 & \textbf{43.7} & \textbf{95.9} & \textbf{97.0} \\
\rowcolor{oursblue}\textbf{Ours} & \textbf{67.1}\smallstd{0.35} & \textbf{95.2}\smallstd{0.04} & \textbf{54.9}\smallstd{0.42} & 58.3\smallstd{0.00} & \textbf{78.1}\smallstd{0.14} & 43.0\smallstd{0.47} & 95.3\smallstd{0.10} & \textbf{97.0}\smallstd{0.00} \\
\midrule\textbf{Llama-Nemotron-8B} & 72.7 & 95.1 & 61.1 & 67.1 & 77.7 & \textbf{46.4} & 95.0 & \textbf{99.0} \\
\rowcolor{oursblue}Ours (training-free) & 77.5 & \textbf{97.0} & 65.5 & 70.1 & 81.9 & 46.2 & \textbf{97.3} & 98.0 \\
\rowcolor{oursblue}\textbf{Ours} & \textbf{77.7}\smallstd{0.06} & 96.5\smallstd{0.04} & \textbf{66.2}\smallstd{0.19} & \textbf{70.3}\smallstd{0.27} & \textbf{82.1}\smallstd{0.06} & 46.2\smallstd{0.08} & \textbf{97.3}\smallstd{0.00} & 98.3\smallstd{0.58} \\
\bottomrule\end{tabularx}\end{table}

\begin{table}[H]\centering
\caption{\textbf{Per-task NDCG@10 for the six additional backbones in Table~\ref{tab:backbone-summary} (\%).} Backbone rows report Original embedding; Ours reports mean $\pm$ SD.}
\fontsize{8.0}{9.0}\selectfont\setlength{\tabcolsep}{1.1pt}\renewcommand{\arraystretch}{0.96}
\begin{tabularx}{\linewidth}{l*{7}{>{\centering\arraybackslash}X}>{\centering\arraybackslash}X}\toprule
Model & LC & LM & RT & ESG & QA & PQ & CQ & MLDR \\
\midrule\textbf{Llama-1B} & 54.2 & 83.1 & 48.0 & 48.4 & 56.2 & 36.2 & 84.4 & 80.8 \\
\rowcolor{oursblue}Ours (training-free) & 59.0 & \textbf{89.1} & 51.2 & 54.1 & 56.6 & 37.3 & 89.6 & 83.4 \\
\rowcolor{oursblue}\textbf{Ours} & \textbf{59.1}\smallstd{0.10} & 88.7\smallstd{0.30} & \textbf{52.1}\smallstd{0.08} & \textbf{54.2}\smallstd{0.33} & \textbf{57.1}\smallstd{0.11} & \textbf{37.7}\smallstd{0.10} & \textbf{89.8}\smallstd{0.06} & \textbf{84.2}\smallstd{0.16} \\
\midrule\textbf{Llama-3B} & 56.3 & 75.4 & 49.2 & 57.0 & 58.2 & 34.6 & 76.8 & 80.1 \\
\rowcolor{oursblue}Ours (training-free) & 62.4 & \textbf{87.5} & \textbf{54.1} & \textbf{64.2} & \textbf{59.1} & \textbf{38.8} & \textbf{87.7} & \textbf{85.1} \\
\rowcolor{oursblue}\textbf{Ours} & \textbf{62.7}\smallstd{0.06} & 86.2\smallstd{0.02} & 53.9\smallstd{0.16} & 63.5\smallstd{0.57} & 59.0\smallstd{0.21} & 37.8\smallstd{0.19} & 87.3\smallstd{0.01} & 83.8\smallstd{0.64} \\
\midrule\textbf{EmbeddingGemma-300M} & 50.4 & 79.8 & 48.4 & 41.3 & 56.9 & 37.0 & 79.1 & 82.1 \\
\rowcolor{oursblue}Ours (training-free) & 61.4 & \textbf{86.7} & 51.4 & \textbf{63.0} & 59.1 & \textbf{40.0} & \textbf{88.1} & 87.1 \\
\rowcolor{oursblue}\textbf{Ours} & \textbf{61.5}\smallstd{0.06} & 86.6\smallstd{0.11} & \textbf{51.9}\smallstd{0.17} & 61.9\smallstd{0.28} & \textbf{59.2}\smallstd{0.04} & 39.7\smallstd{0.15} & 88.0\smallstd{0.04} & \textbf{87.9}\smallstd{0.07} \\
\midrule\textbf{Jina-v5-text-nano} & 45.0 & 77.4 & 41.2 & 49.1 & 48.5 & 32.3 & 78.5 & 80.0 \\
\rowcolor{oursblue}Ours (training-free) & 58.3 & \textbf{85.9} & 49.2 & \textbf{61.4} & 53.2 & 35.7 & 86.7 & 84.8 \\
\rowcolor{oursblue}\textbf{Ours} & \textbf{58.7}\smallstd{0.01} & 85.8\smallstd{0.02} & \textbf{49.5}\smallstd{0.28} & 60.7\smallstd{0.11} & \textbf{53.8}\smallstd{0.23} & \textbf{36.2}\smallstd{0.46} & \textbf{87.1}\smallstd{0.09} & \textbf{85.1}\smallstd{0.22} \\
\midrule\textbf{Qwen3-Embedding-8B} & 36.3 & 77.4 & 41.0 & 44.2 & 47.4 & 30.0 & 79.4 & 77.4 \\
\rowcolor{oursblue}Ours (training-free) & 47.6 & 86.4 & 42.4 & \textbf{46.7} & 50.3 & 34.2 & \textbf{84.4} & 80.7 \\
\rowcolor{oursblue}\textbf{Ours} & \textbf{53.0}\smallstd{0.25} & \textbf{88.5}\smallstd{0.05} & \textbf{44.7}\smallstd{0.20} & 45.5\smallstd{0.05} & \textbf{54.2}\smallstd{0.18} & \textbf{34.5}\smallstd{0.46} & 84.2\smallstd{0.05} & \textbf{82.1}\smallstd{0.21} \\
\midrule\textbf{Llama-Nemotron-8B} & 58.2 & 85.8 & 51.3 & 50.9 & 56.5 & 35.6 & 81.0 & 83.4 \\
\rowcolor{oursblue}Ours (training-free) & 63.4 & \textbf{90.3} & 56.0 & \textbf{60.1} & 60.0 & \textbf{37.8} & 89.5 & 85.2 \\
\rowcolor{oursblue}\textbf{Ours} & \textbf{64.0}\smallstd{0.10} & 90.0\smallstd{0.23} & \textbf{57.0}\smallstd{0.12} & \textbf{60.1}\smallstd{0.66} & \textbf{60.6}\smallstd{0.14} & 37.7\smallstd{0.15} & \textbf{89.6}\smallstd{0.03} & \textbf{85.8}\smallstd{0.17} \\
\bottomrule\end{tabularx}\end{table}

\FloatBarrier
\section{Additional Scoring Analysis}\label{app:ablations}
\noindent\textbf{Motivating example.}
Figure~\ref{fig:motivation} uses the saved training-free Qwen3-Embedding-4B configuration on LMEB LoCoMo, with token budget $k=64$ and equal-weight standardized fusion. We jointly remove the six highlighted query tokens' excess-over-floor contributions across all candidate memories. Query and memory encodings, whole-memory scores, the query-token denominator, and the original local normalization remain fixed. The relevant memory changes from fused rank 1 to rank 32. This intervention removes score contributions, not text, and does not re-encode the inputs or renormalize the remaining local scores. Removing only the contribution of ``progress'' leaves the relevant memory at rank 1. The displayed memories are excerpts, and word groups are used for readability rather than exact tokenizer boundaries. The match between ``progress'' and ``come'' is contextual, within ``how far I've come,'' rather than a claim that the isolated words are synonyms.

\noindent\textbf{Top-10 recovery counts.}
The bars in Figure~\ref{fig:motivation} count queries for which original retrieval has no labelled relevant memory in Top-10 but training-free fusion retrieves at least one. Across all 1,976 saved LMEB LoCoMo queries, including all five question types, there are 607 such recoveries and 84 reverse transitions, giving a net gain of 523 hit queries. The original labels and query set are retained. These are gross query-level counts, not category-normalized rates or Recall@10. They do not imply a net improvement in every category: open-domain queries have 9 recoveries and 10 reverse transitions.

\noindent\textbf{Independent chunking at shorter windows.}
The main Chunk-MaxP baseline uses 512-token chunks with 64-token overlap. On Qwen-0.6B LoCoMo, shorter independent windows of 128, 64, and 32 tokens yield Recall@10 of 44.99--46.87 and NDCG@10 of 33.42--34.23. Original embedding reaches 46.87/34.24 Recall@10/NDCG@10, compared with 67.98/53.42 for our method without query adaptation. In this setting, reducing the window size does not close the gap to token matching and fusion.

\begin{table}[!htbp]\centering
\caption{\textbf{Effects of exact search and missing-score estimation at $k=64$.} Adapted Recall@10/NDCG@10 (\%), averaged over three training seeds.}\label{tab:k-tail-exact}
\small\setlength{\tabcolsep}{5pt}\renewcommand{\arraystretch}{1.0}
\begin{tabular*}{\linewidth}{@{\extracolsep{\fill}\hspace{1.1pt}}llrrr@{\hspace{1.1pt}}}\toprule
Model & Dataset & HNSW + floor & Exact + floor & HNSW + zero \\\midrule
Qwen-0.6B & LoCoMo & 68.69/54.51 & 69.30/55.31 & 61.84/47.04 \\
Qwen-0.6B & LongMemEval & 94.06/87.48 & 94.01/87.50 & 92.46/82.35 \\
Qwen-4B & LoCoMo & 67.83/53.58 & 69.44/55.49 & 60.69/45.60 \\
Qwen-4B & LongMemEval & 95.94/88.15 & 95.96/88.68 & 93.99/85.58 \\
\bottomrule\end{tabular*}
\end{table}

\begin{table}[!htbp]\centering
\caption{\textbf{Scoring controls, Recall@10/NDCG@10 (\%).}}\label{tab:controls}
\fontsize{6.9}{8.3}\selectfont\setlength{\tabcolsep}{1.1pt}\renewcommand{\arraystretch}{1.02}
\begin{tabularx}{\linewidth}{l*{3}{>{\hsize=.93\hsize\centering\arraybackslash}X>{\hsize=1.07\hsize\centering\arraybackslash}X p{3pt}}>{\hsize=.93\hsize\centering\arraybackslash}X>{\hsize=1.07\hsize\centering\arraybackslash}X}\toprule
 & \multicolumn{2}{c}{MaxSim} &  & \multicolumn{2}{c}{MaxSim + Whole} &  & \multicolumn{2}{c}{Sparse} &  & \multicolumn{2}{c}{Sparse + Whole} \\
\cmidrule{2-3}\cmidrule{5-6}\cmidrule{8-9}\cmidrule{11-12}
Task & \multicolumn{1}{c}{Training-free} & \multicolumn{1}{c}{Query-adapted} &  & \multicolumn{1}{c}{Training-free} & \multicolumn{1}{c}{Query-adapted} &  & \multicolumn{1}{c}{Training-free} & \multicolumn{1}{c}{Query-adapted} &  & \multicolumn{1}{c}{Training-free} & \multicolumn{1}{c}{Query-adapted} \\\midrule
\multicolumn{12}{l}{\textit{Qwen-0.6B}} \\
LoCoMo & 64.2/51.3 & 66.3/53.7 &  & 61.6/48.9 & 63.6/50.3 &  & 66.7/52.1 & 67.4/53.3 &  & 68.0/53.4 & 68.7/54.5 \\
LongMemEval & 92.2/82.5 & 92.7/83.2 &  & 93.0/84.3 & 93.2/84.7 &  & 93.0/86.4 & 93.3/86.2 &  & 94.5/87.1 & 94.1/87.5 \\
REALTALK & 53.4/43.8 & 55.9/45.9 &  & 54.7/44.6 & 56.4/45.6 &  & 51.8/41.9 & 53.7/44.3 &  & 55.1/44.9 & 55.6/45.9 \\
ESG & 54.3/37.7 & 57.0/39.7 &  & 56.3/44.9 & 56.3/45.3 &  & 66.3/47.1 & 65.8/49.9 &  & 64.1/52.3 & 65.1/51.3 \\
QASPER & 73.8/47.1 & 75.0/48.4 &  & 75.5/52.1 & 75.9/52.7 &  & 74.0/47.6 & 74.5/48.3 &  & 76.2/51.7 & 76.6/52.5 \\
PeerQA & 43.0/32.2 & 43.5/31.9 &  & 42.9/32.8 & 44.1/32.8 &  & 41.0/32.8 & 42.8/33.2 &  & 44.6/34.0 & 44.3/34.9 \\
CovidQA & 94.4/82.7 & 94.5/83.7 &  & 94.9/82.9 & 95.2/83.0 &  & 94.4/83.1 & 94.6/84.1 &  & 95.4/83.8 & 95.9/84.4 \\
MLDR & 96.0/80.7 & 97.0/82.3 &  & 96.0/81.6 & 96.0/82.0 &  & 96.0/80.6 & 96.0/81.4 &  & 96.0/81.6 & 96.0/81.7 \\
\midrule\multicolumn{12}{l}{\textit{Qwen-4B}} \\
LoCoMo & 54.2/43.3 & 66.0/53.6 &  & 54.9/42.2 & 62.2/48.3 &  & 62.1/49.2 & 66.4/52.7 &  & 63.5/49.7 & 67.8/53.6 \\
LongMemEval & 93.3/84.6 & 93.1/85.1 &  & 92.9/84.7 & 94.4/86.2 &  & 94.7/87.0 & 95.2/87.2 &  & 95.4/87.7 & 95.9/88.2 \\
REALTALK & 48.4/40.5 & 57.6/48.6 &  & 51.6/41.9 & 55.3/45.5 &  & 50.9/41.3 & 55.5/45.7 &  & 52.3/42.7 & 56.6/46.7 \\
ESG & 49.2/31.9 & 53.1/40.4 &  & 55.1/44.2 & 55.1/45.7 &  & 58.7/40.2 & 64.2/48.4 &  & 68.0/49.3 & 70.8/53.3 \\
QASPER & 68.2/42.9 & 76.7/50.4 &  & 75.5/52.0 & 78.1/54.4 &  & 68.6/43.4 & 76.5/50.6 &  & 75.1/51.1 & 78.9/54.4 \\
PeerQA & 39.3/30.0 & 44.5/34.6 &  & 40.1/30.9 & 41.7/33.6 &  & 42.5/32.4 & 46.2/35.0 &  & 44.2/33.7 & 44.0/34.2 \\
CovidQA & 94.6/82.6 & 95.3/83.1 &  & 96.0/84.9 & 96.0/84.6 &  & 95.1/83.1 & 95.8/83.6 &  & 95.7/85.3 & 95.9/85.5 \\
MLDR & 95.0/79.0 & 96.0/81.8 &  & 95.0/80.0 & 96.0/82.5 &  & 94.0/79.1 & 96.0/81.8 &  & 96.0/80.0 & 96.0/82.5 \\
\bottomrule\end{tabularx}\end{table}

\FloatBarrier
\section{Compression and Cost Measurement}\label{app:quantization}\label{app:cost}
Table~\ref{tab:compression} reports the quality--storage trade-off across tasks, while Table~\ref{tab:cost-qwen-complete} adds the latency comparison for both Qwen backbones.

\noindent\textbf{Measurement settings.}
Storage and latency use the 568-candidate LoCoMo bank with median candidate-text volume; quality uses all 1,976 LoCoMo queries. Adapted Qwen configurations use the source-selected checkpoint from seed 31070. Latency is the median over eight fixed queries and five warm repeats, including query encoding and ranking. Methods run separately with one CPU math thread on a shared server; BM25 and XTR run on CPU, while other neural methods use GPU acceleration. Storage counts the index, token-to-memory mappings, pooled vectors, candidate IDs, and deployed adapter, excluding encoder weights, source text, and caches. RP128 uses the same fixed projection for memory and query states without subsequent normalization.
\begin{table}[H]\centering
\caption{\textbf{Retrieval quality and storage under compression.} Quality is Recall@10/NDCG@10, evaluated on all queries and averaged over three query adapters; storage is measured on a fixed representative candidate bank per task.}\label{tab:compression}
\fontsize{7.5}{9.0}\selectfont\setlength{\tabcolsep}{1.1pt}\renewcommand{\arraystretch}{1.03}
\begin{tabularx}{\linewidth}{l*{5}{>{\hsize=1.15\hsize\centering\arraybackslash}X>{\hsize=.85\hsize\centering\arraybackslash}X p{1.8pt}}>{\hsize=1.15\hsize\centering\arraybackslash}X>{\hsize=.85\hsize\centering\arraybackslash}X}\toprule
 & \multicolumn{2}{c}{Full} &  & \multicolumn{2}{c}{SQ4} &  & \multicolumn{2}{c}{RP128+SQ4} &  & \multicolumn{2}{c}{Pool 4} &  & \multicolumn{2}{c}{Pool 8} &  & \multicolumn{2}{c}{Pool 16} \\
\cmidrule{2-3}\cmidrule{5-6}\cmidrule{8-9}\cmidrule{11-12}\cmidrule{14-15}\cmidrule{17-18}
Task & R/N & \multicolumn{1}{c}{MiB} &  & R/N & \multicolumn{1}{c}{MiB} &  & R/N & \multicolumn{1}{c}{MiB} &  & R/N & \multicolumn{1}{c}{MiB} &  & R/N & \multicolumn{1}{c}{MiB} &  & R/N & \multicolumn{1}{c}{MiB} \\
\midrule
\multicolumn{18}{l}{\textit{Qwen3-Embedding-0.6B}} \\
LoCoMo & 68.7/54.5 & 153.9 &  & 68.3/54.3 & 26.2 &  & 66.8/52.0 & 10.7 &  & 65.9/49.5 & 11.7 &  & 67.3/50.9 & 20.9 &  & 66.2/50.7 & 39.3 \\
LongMemEval & 94.1/87.5 & 597.0 &  & 94.6/87.6 & 95.1 &  & 94.9/86.5 & 32.9 &  & 90.1/76.6 & 9.8 &  & 92.0/80.4 & 17.6 &  & 92.9/82.8 & 33.0 \\
REALTALK & 55.6/45.9 & 173.1 &  & 55.1/45.4 & 29.5 &  & 54.6/44.8 & 12.0 &  & 53.3/42.0 & 13.6 &  & 51.6/42.2 & 24.3 &  & 51.4/41.1 & 45.8 \\
ESG & 65.1/51.3 & 86.3 &  & 68.3/52.5 & 14.1 &  & 59.4/47.4 & 5.5 &  & 66.7/41.8 & 2.3 &  & 64.6/42.9 & 4.0 &  & 57.0/40.2 & 7.3 \\
QASPER & 76.6/52.5 & 19.9 &  & 76.7/52.6 & 3.5 &  & 74.7/51.2 & 1.9 &  & 72.3/48.0 & 1.3 &  & 72.0/47.9 & 2.1 &  & 71.8/47.6 & 3.7 \\
PeerQA & 44.3/34.9 & 34.6 &  & 44.6/35.2 & 6.4 &  & 44.6/33.2 & 3.4 &  & 41.0/30.9 & 5.1 &  & 41.8/30.9 & 8.8 &  & 40.3/29.8 & 15.8 \\
CovidQA & 95.9/84.4 & 22.9 &  & 95.9/84.2 & 3.9 &  & 94.7/82.0 & 2.0 &  & 91.9/74.1 & 0.9 &  & 92.5/74.5 & 1.4 &  & 93.0/75.0 & 2.5 \\
MLDR & 96.0/81.7 & 9.7 &  & 96.0/81.7 & 1.8 &  & 97.0/82.9 & 1.3 &  & 98.0/80.8 & 0.5 &  & 98.0/78.0 & 0.7 &  & 96.3/78.3 & 1.1 \\
\midrule\multicolumn{18}{l}{\textit{Qwen3-Embedding-4B}} \\
LoCoMo & 67.8/53.6 & 376.6 &  & 68.6/54.2 & 57.3 &  & 63.3/49.5 & 15.2 &  & 62.8/46.7 & 28.7 &  & 65.3/49.1 & 51.2 &  & 64.2/49.1 & 96.3 \\
LongMemEval & 95.9/88.2 & 1460.4 &  & 96.2/88.6 & 205.8 &  & 94.5/85.8 & 36.8 &  & 92.8/80.6 & 24.2 &  & 94.0/82.9 & 43.0 &  & 95.1/84.1 & 80.8 \\
REALTALK & 56.6/46.7 & 423.7 &  & 57.4/47.6 & 64.5 &  & 52.1/43.2 & 17.0 &  & 49.5/39.5 & 33.4 &  & 48.8/39.7 & 59.6 &  & 47.6/38.1 & 112.1 \\
ESG & 70.8/53.3 & 211.2 &  & 71.0/53.4 & 30.5 &  & 65.4/48.0 & 7.3 &  & 66.1/42.5 & 5.7 &  & 62.2/41.2 & 9.8 &  & 65.5/42.4 & 18.0 \\
QASPER & 78.9/54.4 & 48.6 &  & 78.8/54.4 & 7.7 &  & 76.5/52.6 & 3.4 &  & 74.7/50.1 & 3.1 &  & 74.8/49.5 & 5.1 &  & 73.6/48.1 & 9.0 \\
PeerQA & 44.0/34.2 & 84.7 &  & 43.9/34.3 & 14.3 &  & 43.5/34.0 & 6.0 &  & 36.3/25.9 & 12.6 &  & 39.2/26.9 & 21.5 &  & 38.3/27.4 & 38.8 \\
CovidQA & 95.9/85.5 & 55.9 &  & 95.9/85.5 & 8.5 &  & 95.6/83.5 & 3.3 &  & 92.6/75.7 & 2.2 &  & 93.1/75.5 & 3.5 &  & 93.5/75.5 & 6.0 \\
MLDR & 96.0/82.5 & 23.7 &  & 96.0/82.8 & 3.9 &  & 97.0/81.1 & 2.5 &  & 99.0/78.9 & 1.3 &  & 98.0/77.9 & 1.8 &  & 99.0/78.9 & 2.8 \\
\bottomrule\end{tabularx}
\end{table}

\begin{table}[H]\centering
\caption{\textbf{Retrieval quality, storage, and latency on LoCoMo.}}\label{tab:cost-qwen-complete}
\fontsize{8.0}{9.4}\selectfont\setlength{\tabcolsep}{1.1pt}\renewcommand{\arraystretch}{1.04}
\begin{tabularx}{\linewidth}{l*{3}{>{\centering\arraybackslash}X}>{\centering\arraybackslash}X}\toprule
Method & Recall@10 $\uparrow$ & NDCG@10 $\uparrow$ & Size (MiB) $\downarrow$ & Query (ms) $\downarrow$ \\
\midrule\multicolumn{5}{l}{\textit{Qwen3-Embedding-0.6B}} \\
Original embedding & 46.87 & 34.24 & 2.23 & 16.66 \\
Search-Adaptor & 49.08 & 35.62 & 4.24 & 16.83 \\
\rowcolor{oursblue}Ours (Full) & 68.53 & 54.45 & 153.93 & 29.21 \\
\rowcolor{oursblue}Ours (SQ4) & 68.37 & 54.31 & 26.17 & 21.68 \\
\rowcolor{oursblue}Ours (RP128+SQ4) & 66.69 & 51.92 & 10.70 & 20.69 \\
\midrule\multicolumn{5}{l}{\textit{Qwen3-Embedding-4B}} \\
Original embedding & 41.60 & 29.95 & 5.56 & 22.71 \\
Search-Adaptor & 44.79 & 32.31 & 10.58 & 24.91 \\
\rowcolor{oursblue}Ours (Full) & 67.93 & 53.50 & 376.65 & 35.12 \\
\rowcolor{oursblue}Ours (SQ4) & 68.51 & 54.11 & 57.27 & 33.15 \\
\rowcolor{oursblue}Ours (RP128+SQ4) & 63.09 & 49.33 & 15.15 & 29.31 \\
\bottomrule\end{tabularx}
\end{table}

\FloatBarrier
\section{Downstream Question-Answering Details}\label{app:qa}
\noindent\textbf{Evaluation protocol.}
We use Qwen3-30B-A3B-Instruct-2507 as the answering model and GPT-4o-mini-2024-07-18 as the judge. The main comparison retrieves $K=10$ memories; the budget analysis uses $K\in\{4,8,10,16\}$. Memory construction and task-specific prompts follow EvoEmbedding \citep{nie2026evo}. One API-rejected input for Qwen-8B Ours on LongMemEval was answered locally with the same named reader's BF16 weights and greedy decoding, then scored by the same API judge; all other answers used the API reader.

\noindent\textbf{Metrics.}
Native retrieval uses Recall and binary-relevance NDCG on 1,533 LoCoMo questions with resolved evidence labels and 470 non-abstention LongMemEval questions. Question answering retains all 1,540 and 500 questions, respectively, with question-weighted overall accuracy and F1. LoCoMo F1 follows its category-aware evaluation code. LongMemEval F1 is a supplemental lexical metric applying the same normalization and stemming to the full predicted and reference answers, including abstention questions.

\end{document}